\documentclass[aps,prd,twocolumn,superscriptaddress,
floatfix,nofootinbib]{revtex4-1}

\pdfoutput=1

\usepackage[dotinlabels]{titletoc}
\usepackage{titlesec}

\usepackage{xcolor}
\usepackage{mathrsfs}
\usepackage{amsmath}
\usepackage{amssymb}
\usepackage{bm}
\usepackage{amsfonts}
\usepackage{subfigure}
\usepackage{feynmp}
\usepackage{extarrows}
\usepackage{slashed}
\usepackage{graphicx}
\usepackage{dcolumn}
\usepackage{verbatim}
\usepackage{here}
\usepackage{multirow} 
\allowdisplaybreaks[4]
\usepackage{indentfirst}
\usepackage{isodateo}
\usepackage[low-sup]{subdepth}

\definecolor{gesfpurple}{rgb}{0.47,0.19,0.42}

\definecolor{gesflanse}{rgb}{0.00,0.50,0.50}

\definecolor{gesfblue}{rgb}{0.08,0.42,0.76}

\definecolor{gesfred}{rgb}{1,0,0}

\definecolor{gesfwhite}{rgb}{1,1,1}

\definecolor{gesfblack}{rgb}{0,0,0}

\usepackage{isodateo}
\usepackage[CJKbookmarks=true, bookmarksnumbered=true,bookmarksopen=true,]{hyperref}
\hypersetup{colorlinks,%
              linkcolor=blue,
              citecolor=blue,
              urlcolor=blue}

\graphicspath{{figs/}}

\newcommand{\be}{\begin{equation}}
\newcommand{\ee}{\end{equation}}
\newcommand{\bea}{\begin{eqnarray}}
\newcommand{\eea}{\end{eqnarray}}

\def\ede{\end{equation}}
\def\bga{\begin{aligned}}
\def\eda{\end{aligned}}
\newcommand{\beq}{\begin{equation}}
\newcommand{\eeq}{\end{equation}}
\newcommand{\bq}{\begin{equation}}
\newcommand{\eq}{\end{equation}}
\newcommand{\ba}{\begin{array}}
\newcommand{\ea}{\end{array}}
\newcommand{\beqa}{\begin{eqnarray}}
\newcommand{\eeqa}{\end{eqnarray}}
\newcommand{\beqs}{\begin{subequations}}
\newcommand{\eeqs}{\end{subequations}}

\newcommand{\fr}[2]{\mbox{$\frac{\,{#1}\,}{#2}$}}
\def\nn{\nonumber}

\def\({\left(}
\def\){\right)}

\def\End{\end{document}}

\def\d{\text{d}}
\def\ii{{\tt i}}
\def\over{\overline}

\def\be{\beta}

\def\Dm{\Delta{m}}
\def\mX{m_{\!X}^{}}

\def\mAP{m_{\!A'}^{}}

\def\End{\end{document}}

\begin{document}

\title{GeV Scale Inelastic Dark Matter with Dark Photon Mediator\\
	via Direct Detection and Cosmological/Laboratory Constraints}

\author{{Hong-Jian He}}
\email[]{hjhe@sjtu.edu.cn; hjhe@tsinghua.edu.cn}
\affiliation{%
Tsung-Dao Lee Institute $\&$ School of Physics and Astronomy,\\
Key Laboratory for Particle Astrophysics and Cosmology (MOE),\\
Shanghai Key Laboratory for Particle Physics and Cosmology,\\
Shanghai Jiao Tong University, Shanghai 200240, China}
\affiliation{%
Institute of Modern Physics $\&$ Physics Department,
Tsinghua University, Beijing 100084, China}
\affiliation{%
Center for High Energy Physics, Peking University, Beijing 100871, China}

\author{{Yu-Chen Wang$^{1,2,}$}}
\email[]{wangyc21@sjtu.edu.cn}

\author{{Jiaming Zheng}$^1$}
\email[]{zhengjm3@sjtu.edu.cn}

\begin{abstract} 
\noindent 	
We propose a new candidate of GeV scale inelastic dark matter (DM).
Our construction has an anomaly-free $U(1)_X^{}$ gauge group with dark photon mediator, and can realize either scalar or fermionic inelastic DM. It is highly predictive and testable. 
We study the scattering rate of light inelastic DM with electrons in the XENON1T experiment and with nuclei in the XENON1T, CRESST-III, CDEX-1B and DarkSide-50 experiments. We resolve the recent XENON1T anomaly via electron recoil detection. Combining the XENON1T constraints from both electron recoils and nuclear recoils (including Migdal effect), we predict the inelastic DM mass $\lesssim\!1.5\,$GeV.
We further analyze the bounds by the DM relic abundance, the lifetime of the heavier DM component, and laboratory constraints, from which we identify the viable parameter space for the future probe. This provides an important benchmark for the theories and experimental tests of GeV scale inelastic DM.
\\[2mm]
Phys.\ Rev.\ D\,104 (2021) 115033, no.11 [arXiv:2012.05891].
\end{abstract}

\maketitle	

\section{\hspace*{-2.5mm}Introduction}
\label{sec:1}
\vspace*{-3mm}

Searching for GeV scale light dark matter (DM) particles is a very challenging
task because the conventional direct detection via DM-nucleon recoil becomes difficult
for DM masses $\lesssim\!5$\,GeV.
Hence, measuring the DM-electron recoil spectrum has provided an important means
for light DM direct detection.
The XENON collaboration\,\cite{Aprile:2020yad} recently announced
a $3.5\sigma$ excess of events with low electron recoil energy
in its Science\,Run-I\,\cite{Aprile:2020tmw}.
The XENON1T detector recorded 285 events for the recoil energy
$E_R^{}\!=\!(1\!-\!7)\,{\rm keV}$,
among which the expected background events are $\,232\pm 15$ \cite{Aprile:2020tmw}.
This excess centers around $E_R^{}\!=\!(2\!-\!3.5)\,{\rm keV}$.
Lately, the PandaX-II collaboration also reported an independent DM search
by measuring the low energy electron recoil spectrum with robust estimates of backgrounds\,\cite{PandaX2}.
It is consistent with the XENON1T measurement\,\cite{Aprile:2020tmw}
although its sensitivity is not yet enough to either confirm
or exclude the DM interpretation of the XENON1T anomaly.
There have been possible explanations for this excess in the literature,
including unexpected tritium background\,\cite{Aprile:2020tmw,Tritium} and
various new physics models\,\cite{Strumia,Inelastic,He:2020wjs,others}.

\vspace*{1mm}

One attractive resolution of the XENON1T anomaly is the exothermic inelastic scattering\,\cite{Inelastic,He:2020wjs}
between the DM and electron.
In this scenario, the DM consists of two components
$(X,\,X')$ with a small mass-splitting
$\,\Delta m_{\!X}\equiv m_{X'}^{}\!-\!m_X^{}$,
which is around the anomalous recoil energy region
$(2\!-\!3)\,{\rm keV}$ of XENON1T.
The heavier DM component $X'$ is cosmologically stable
because its decay to the lighter DM component $X$ is highly suppressed
by the small mass-splitting $\Delta m_{\!X}$\,.
Inside the xenon detector, $X'$ scatters inelastically
with the xenon electron and de-excite to $X$.%
\footnote{%
This differs in an essential way from the well-studied endothermic inelastic DM scattering in the literature\,\cite{Gu:2018kmv}
which cannot explain XENON1T anomaly and is irrelevant to the
our current work.}
The DM mass-splitting manifests itself as a peak
in the electron recoil energy spectrum.
On the other hand, the recoil energy from the elastic DM-electron scattering
is too small to be detectable in XENON1T unless the DM particles are
very fast moving\,\cite{Strumia}.
To generate the XENON1T anomaly,
we can estimate the required normalized cross section
of the DM-electron inelastic scattering,
$\,\bar\sigma_e^{}/m_X^{} \!\approx\! 5.7\!\times\! 10^{-44} {\rm cm^2/GeV}$,\,
by considering the DM $(X,\,X')$ of equal amount
and the DM-electron interaction of
scalar or vector type\,\cite{He:2020wjs}.
Here we define
$\,\bar\sigma_e^{}\!\equiv\! \sigma_{Xe}^{}(|{\bf q}|\!=\!0)$\,
as the scattering cross section of
$\,X'e^-\!\!\to X e^-$\,
in the zero-momentum-exchange limit $|{\bf q}| \!=\! 0$\,.

\vspace*{1mm}

In the literature, the inelastic scatterings mostly arise from
exchanging a light dark photon with couplings to leptons by assuming
a tiny kinetic mixing between the dark photon
and the standard model (SM) photon\,\cite{Inelastic}.
These models require a large hierarchy 
between the dark-photon-lepton coupling
and the dark-photon-DM coupling.
So the dark photons could be hidden from various collider searches
by tuning the kinetic mixing of the dark photon with photon
down to $O(10^{-3}\!\!-\!10^{-6})$.
The origin of such tiny kinetic mixing remains obscure\,\cite{Gherghetta:2019coi},
so we will not pursue this for the present study.

\vspace*{1mm}

In this work, we propose a new candidate of GeV scale inelastic DM.
We realize this in an anomaly-free and renormalizable model,
in which the DM and the right-handed first family fermions join the interactions of a dark $U(1)_X$ gauge group.
We achieve the desired $O(\text{keV})$ mass-splitting of the
inelastic DM by a scalar seesaw mechanism without fine-tuning.
We show that this minimal model is viable 
for the inelastic DM of mass
{ $\lesssim\!1.5$\,GeV,}
which can provide the intriguing anomaly
in the XENON1T electron recoil spectrum\,\cite{Aprile:2020tmw}
and be consistent with the DM-nucleon recoil detections 
(of low threshold)
by XENON1T\,\cite{Aprile:2019jmx}, CRESST-III\,\cite{Abdelhameed:2019hmk},
CDEX-1B\,\cite{CDEX-1B}, and DarkSide-50\,\cite{DarkS} experiments.
For the dark photon mediator with mass $<\!2\hspace*{1pt}\mX$,
this model can provide the observed DM relic abundance
and ensure the heavier DM component cosmologically stable.
We further derive non-trivial bounds from the existing laboratory measurements, including the electroweak precision tests 
and collider searches. 
We also discuss the possible future experimental probes.

\vspace*{1mm}

This paper is organized as follows.
We construct our model in Sec.\,\ref{sec:2}.
Then, in Sec.\,\ref{sec:DD_Event_Rates}
we analyze the DM-electron and DM-nucleon recoil signals
in various direct detection experiments.
In Sec.\,\ref{sec:CosmoConstraints}, we study the cosmological constraints on our model,
including the $X'$ lifetime and the DM relic abundance.
In Sec.\,\ref{sec:LabConstraints}, we study other laboratory constraints,
including the electroweak precision tests and the collider searches.
Finally, we conclude in Sec.\,\ref{sec:Conclusion}.  In Appendix\,\ref{app:A}, we propose an improved treatment of the Migdal effect bound on the inelastic DM.
Appendix\,\ref{app:B} presents our analysis on the Higgs sector 
of this model.

\vspace*{-2mm}
\section{\hspace*{-2.5mm}Inelastic DM with Dark Photon Mediator} 
\label{sec:2}
\vspace*{-2mm}

To realize the DM-electron interaction,
we construct a minimal extension of the SM by a dark $U(1)_X$ gauge group
under which both the DM and the right-handed first family fermions
are charged. 
We also include three right-handed Majorana neutrinos $\nu_{Rj}^{}$
($j\!=\!1,2,3$).
We denote the $U(1)_X$ gauge boson (dark photon) by $A'_\mu$\,.
The Higgs sector consists of two Higgs doublets plus three singlet scalars
$S$, $S'$ and $\phi$\,,\, charged under $U(1)_X$.
The electroweak symmetry breaking is realized by two
Higgs doublets $H_1^{}$ and $H_2^{}$ with vacuum expectation values (VEVs)
$\langle H_j^{} \rangle\!=\!(0,v_j^{})^{T}$ and their combined VEV
$\,v_h^{}\!=\! \sqrt{v_1^2\!+\!v_2^2\,} \!\simeq\! 174$\,GeV.
We will set $v_1^2\!\ll\! v_2^2$\,,\, so
the observed Higgs boson\,($125$GeV) is mostly made of the CP even neutral
component of $H_2^{}$\,.
The dark $U(1)_X$ gauge group is mainly broken
by the VEVs of the singlet scalars $S$ and $S'$,
whose VEVs
$\,\left<S\right>\!=v_S^{}$ and
$\,\left<S'\right>\!=v_S'$\, are of $O(100\text{GeV})$.
Our model sets the first family fermions charged under $U(1)_X$,
and the second and third family fermions as $U(1)_X$ singlets.
In the following, we will study the case of scalar DM
$\,\widehat{X}\!=\!(X\!+\!\ii X')/\!\sqrt{2}$\,
and the case of fermionic DM
$\,\widehat{\chi}\!=\!(\chi_1,\,{\chi_2}^{\dag})^T$, respectively.
In Table\,\ref{tab:1}, we present the particle content and
charge assignments of our model for the dark sector,
the Higgs sector, and the first family fermions.

We note that in the lepton sector, 
only the right-handed $(e_R^{},\nu_{R1}^{})$ are charged under $U(1)_X$ to maintain the stability of the heavier DM component.  
The $U(1)_X$ charge assignments of the first family SM fermions 
(including the right-handed neutrino) are then uniquely determined by
the anomaly cancellation.

\vspace*{-2mm}
\subsection{\hspace*{-2.5mm}Inelastic Scalar DM with Dark \boldmath{$U(1)_X^{}$}}
\label{sec:SDM}
\label{sec:2.1}
\vspace*{-2mm}

The DM particles $(X,\,X')$ form a complex scalar
$\widehat{X}\!=\!(X\!+\!\ii X')/\!\sqrt{2}$\,
with $U(1)_X$ charge $q_{\widehat{X}}^{}$\,.
As we will show, the spontaneous breaking of $U(1)_X$ will generate
the desired mass-splitting $\Delta m_{\!X}$ between $X$ and $X'$.
Our model sets the left-handed fermion doublets as $U(1)_X$ singlets.
This forbids the decay channel $\,X'\!\!\to\! X \bar\nu \nu$\,,
and thus ensures that the current DM relic abundance consists of
about equal amounts of $(X,\,X')$ so far.
The anomaly cancellation conditions then uniquely determine the
$U(1)_X$ charges of the first family fermions up to an overall normalization factor.
The flavor non-universality of $U(1)_X$ ensures $H_2^{}$
as $U(1)_X$ singlet, so the $A'$-$Z$ mixing is suppressed by
$v_1^2/v_h^2\ll\! 1$\,
and thus experimentally viable,
as will be shown in Sec.\,\ref{sec:CosmoConstraints}.

\vspace*{1mm}

We write down the relevant Lagrangian terms of the DM sector as follows:
\begin{eqnarray}
\Delta {\cal L}_\text{DM} &\!\supset\!&
|D^\mu\! \widehat{X}|^2 \!- m_{\widehat{X}}^2|\widehat{X}|^2
-\lambda_X^{}|\widehat{X}|^4
\nn\\[-2.4mm]
\label{eq:L-DM}
\\[-2.4mm]
&& -\big(\lambda_{{X}^{}\!\phi}^{}\!\widehat{X}^2\!\phi^2
\!+\! \text{h.c.}\big)
\!-\!\sum_i\! \lambda_{X\psi_i^{}}^{}|\widehat{X}|^2|\psi_i^{}|^2,
\hspace*{8mm}
\nn
\end{eqnarray}
\\[-4mm]
where the scalar fields
$\,\psi_i^{}\!=\!H_1^{},H_2^{},S,S',\phi\,$.
According to Table\,\ref{tab:1},
$H_1^{}$ only couples to the first family fermions,
while $H_2^{}$ interacts only with the second and third families
of fermions.
Thus, we can write the Lagrangian including the relevant Yukawa terms with
$\nu_{Rj}^{}$ and the relevant potential terms with scalar singlets:
\vspace*{-1.5mm}
\begin{eqnarray}
\hspace*{-4mm}
\Delta {\cal L} &\!\supset\!&
 \bar{e}_{R}^{}\ii\slashed{D} e_{R}^{}
\!-\! \sum_{i=1}^3\! \Big(\! y^\nu_{i1}  \bar{L}_i \tilde{H}_1^{}\nu_{R1}^{}
\!+\!\!\sum_{j=2}^3\!y^\nu_{ij}  \bar{L}_i \tilde{H}_2^{}\nu_{Rj}^{}
\!+\! \text{h.c.}\!\Big) \hspace*{-4mm}
\nn
\\[-2mm]
&& -\fr{1}{2}\!\Big( y_S^{} \nu_{R1}^T S \nu_{R1}^{}
+\!\sum_{i,j=2}^{3} M_{Rij}^{}\nu_{Ri}^T\nu_{Rj}^{}
\!+ \text{h.c.}\Big)
~~~~~
\nonumber\\
&&
+M_{S}^2|S|^2
\!+M_{S'}^2|S'|^2 \!+ \!(M_{12}'H_1^\dagger H_2 S'\!+\text{h.c.})
\nonumber
\\[1mm]
&&
\!-M_{\phi}^2|\phi|^2
+ \!\big(\lambda_{S\phi}S^3 \phi^*
+\text{h.c.}\big).
\label{eq:DL}
\end{eqnarray}
We note that the cubic term
$\,M_{12}'H_1^\dagger H_2 S'$\,
can ensure the pseudoscalars to be massive.
In Eq.\eqref{eq:DL}, the squared masses $(M_{S}^2,\,M_{S'}^2,\,M_{\phi}^2)$
are all positive, so $S$ and $S'$ acquire VEVs from their potentials directly,
whereas $\phi$ can only obtain a small VEV induced from
$\left<S\right>$ and $\left<S'\right>$.
In our model, the scalar potential holds CP symmetry,
under which all the scalar couplings and VEVs are real.
Eq.\eqref{eq:DL} shows that
the singlet $S$ and the right-handed neutrino $\nu_{R1}^{}$ form a Yukawa interaction
which generates a weak scale Majorana mass
$\,M_{R1}^{}\!\!=\!y_{S1}^{}v_S^{}\,$ for $\nu_{R1}^{}$.
The second and third family right-handed neutrinos are
$U(1)_X$ singlets, so they form Majorana mass terms directly.
Thus, the light neutrino masses
are generated by the type-I seesaw mechanism.

\tabcolsep 1.4pt
\begin{table*}[t]
\renewcommand{\arraystretch}{1.5} 
\begin{tabular}{c||ccc|ccc||c|c|c|c|c||c|c}
			\hline\hline
			Group  & $Q_{L1}^{}$ & $u_{R}$ & $d_{R}^{}$ & $L_1^{}$ & $e_{R}^{}$
			& $\nu_{R1}^{}$ & $H_1^{}$ & $H_2^{}$ & $S$ & $S'$
			& $\phi$ &
			\,$\widehat{X}$\, & ~$\widehat{\chi}$~  \\
			\hline\hline
			$~SU(2)_L^{}~$
			& $\bf 2$ & $\bf 1$ & $\bf 1$ & $\bf 2$ & $\bf 1$
			& $\bf 1$ & $\bf 2$ & $\bf 2$ & $\bf 1$ &  $\bf 1$& $\bf 1$ & $\bf 1$ & $\bf 1$
			\\
			\hline
			$U(1)_Y^{}$ & $\frac{1}{6}$ & $\frac{2}{3}$ & $-\frac{1}{3} $
			& $-\frac{1}{2}$ & $-1$ & $0$ & $\frac{1}{2}$
			& $\frac{1}{2}$ & $0$
			& $0$ & $0$ & $0$ & $0$ \\
			\hline
			$U(1)_X^{}$ & $0$ & $\frac{1}{2}$ & $-\frac{1}{2}$ & $0$ & $-\frac{1}{2}$
			& $\frac{1}{2}$ & $\frac{1}{2}$ & $0$ & $-1$ & $\frac{1}{2}$
			& $-3$ & $3$ & $\frac{3}{2}$  \\
			\hline
			$\mathbb{Z}_2^{}$
			& $+$ & $+$ & $+$ & $+$ & $+$ & $+$ & $+$ & $+$ & $+$ & $+$ & $+$ & $-$ & $-$
			\\
\hline\hline
\end{tabular}
\renewcommand{\arraystretch}{1} 
\vspace*{-1mm}
\caption{\small Particle content and group assignments of our model.
Here $\,Q_{L_1}^{}\!$ ($L_1^{}$) denotes the left-handed weak doublet
of quarks (leptons) in the 1st family of the SM, while the (2nd, 3rd)
families of the SM fermions are $U(1)_X^{}$ singlets and 
$\mathbb{Z}_2^{}$ even. 
The second column from right defines the scalar inelastic DM $\widehat{X}$,
and the last column defines the fermionic inelastic DM
$\widehat{\chi}$ as another setup. 
}
\label{tab:OVR_UV}
\label{tab:1}
\vspace*{-2mm}
\end{table*}

From the Lagrangian \eqref{eq:L-DM}, we see that
the DM mass is determined by the DM quadratic mass term and the DM couplings
to $|H_i|^2$, $|S|^2$ and $|S'|^2$. The mass-splitting between the two
DM components is determined by the unique quartic interaction
$\widehat{X}^2\phi^2$, where
the singlet VEV $\langle \phi\rangle$ is naturally small as generated
by a scalar seesaw%
\footnote{%
The realization of this scalar seesaw mechanism requires the presence of a $U(1)$ symmetry. It is truly attractive and economical to 
identify it as a gauge symmetry $U(1)_X$  
whose gauge boson serves as the portal (mediator) 
between the dark and visible sectors.}
from the potential terms in the last line of Eq.\eqref{eq:DL}.
This is because $\phi$ is much heavier than all the other scalars and leads to
$\,v_\phi^{}\equiv\langle \phi \rangle\!
 \simeq \lambda_{S\phi}^{}v_S^3/M_{\phi}^2$\,.
Thus, we derive the $(X,\,X')$ mass-splitting:
\beqa
\frac{\,m_{X'}^{}\!-\!m_{X}^{}\,}{m_X^{}}
\simeq
\frac{\,\lambda_{{X}\!\phi}v_{\phi}^2\,}{m_X^2}\,
=
\frac{\,2\lambda_{{X}\!\phi}\lambda_{S\phi}^2 v_S^6\,}
     {m_X^2 M_{\phi}^4} .
\label{eq:mX'-mX}
\eeqa
Hence, to realize the desired $O(\text{keV})$ mass-splitting
for explaining the XENON1T anomaly, we can choose the sample inputs
without fine-tuning,
$\,\lambda_{{X}\phi}^{},\lambda_{S\phi}\!=\! O(0.01)$,
$\,v_S^{}\!=\!O(100)$\,GeV, $\,m_X^{}\!= O(\text{GeV})$,
and $M^{}_\phi\!=\!O(\text{TeV})$. This gives $\,v_\phi^{}\!=\!O(10)\,$MeV
and $\,\Delta m_X^{}=O(\text{keV})$.

\vspace*{1mm}

Since the Higgs doublet $H_1^{}$ carries charges of both $U(1)_Y$ and $U(1)_X$,
its VEV induces mass-mixing between their gauge bosons.
Denoting the neutral gauge bosons of
$SU(2)_L$, $U(1)_Y$ and $U(1)_X$ as $(W^3_\mu, B_\mu^{}, \mathcal{X}_\mu^{})$,
we derive their mass-eigenstates $(Z_\mu^{}, A_\mu^{}, A'_\mu)$
to the leading order of the gauge coupling $\,g_X^{}\!\ll\! 1\,$
and VEV ratio $\,v_1^2/v_h^2\!\ll\! 1$\,,
with the mass-eigenvalues,
\begin{eqnarray}
m_A^2 &\!\!=\!& 0 \,,
\nn\\[-1mm]
m_{A'}^2 &\!\!\simeq\!& 2 g_X^2 \!\(\!v_S^2+\frac{1}{4}v_{S'}^2\!\) \!,
\label{eq:boson_masses}
\\[-1mm]
M_{Z}^2 &\!\!\simeq\!& \frac{1}{2}\!\left( g^2\!+ {g'}^2\!
+ g_X^2\frac{v_1^4}{ v_h^4} \right)\! v_h^2 \,,
\hspace*{10mm}
\nn
\end{eqnarray}
and their leading order mixing matrix,
\beqa
\hspace*{-5mm}
\begin{pmatrix}
\!A'_\mu \\[1mm]
A_\mu^{} \\[1mm]
Z_\mu^{}\!
\end{pmatrix}
\!=\!
\begin{pmatrix}
1 & \frac{-g'g_X^{}}{\,g^2+{g'}^2\,}\frac{v_1^2}{v_h^2}
& \frac{g g_X^{}}{\,g^2+{g'}^2\,}\frac{v_1^2}{v_h^2}
\\
0 & \frac{g}{\sqrt{\!g^2+{g'}^2}\,}
& \frac{g'}{\sqrt{g^2+{g'}^2}\,}
\\
\!\frac{-g_X^{}}{\sqrt{g^2+{g'}^2}\,}
\frac{v_1^2}{v_h^2} & \frac{-g'}{\sqrt{g^2+{g'}^2}\,}
&
\frac{g}{\sqrt{g^2+{g'}^2}\,}
\end{pmatrix}
\!\!\!
\begin{pmatrix}
\!\mathcal{X}_\mu^{} \\[1mm]
B_\mu^{} \\[1mm]
W^3_\mu \!
\end{pmatrix}
\!\!. 
\label{eq:boson_mixing}
\eeqa

\vspace*{-2mm}
\subsection{\hspace*{-2.5mm}Inelastic Fermionic DM with Dark \boldmath{$U(1)_X^{}$}}
\label{sec:2.2}
\label{sec:FDM}
\vspace*{-2mm}

The mechanism of realizing $O(\text{keV})$
mass-splitting for the scalar inelastic DM
in Sec.\,\ref{sec:SDM} can be extended to the case of inelastic fermionic DM.
In this subsection, we present a construction of inelastic fermionic DM.
In this model, the charge assignments (Table\,\ref{tab:1})
and the scalar potential \eqref{eq:DL} remain the same as before,
except the fermionic DM $\widehat{\chi}$ has a different $U(1)_X$
charge as shown in the last column of Table\,\ref{tab:1}.
The fermionic DM contains two Weyl spinors $\chi_1$ and $\chi_2$,
with opposite $U(1)_X$ charges
$\,q_{\chi_1}^{}\!\!=\!-q_{\chi_2}^{}\!\!=\!\fr{3}{2}$\,.
They can form a Dirac spinor
$\,\widehat{\chi}\!=\!(\chi_1,\, {\chi}_2^\dag )^T$,
which is vector-like under $U(1)_X$.
Thus, the dark sector contains the following gauge-invariant
Lagrangian terms:
\begin{eqnarray}
\hspace*{-5mm}
\Delta{\cal L} &\!\supset\!&
\chi^\dag_1\ii \bar\sigma^\mu \!D_\mu^{} \chi_1
+
{\chi_2}^\dag\ii \bar\sigma^\mu \!D_\mu^{} \chi_2
-(m_{\widehat{\chi}}^{}\chi_1\chi_2 \!+\text{h.c.})
\nn\\
\hspace*{-5mm}
&&
+\( y_{\phi\chi_1}^{} \chi_1\chi_1\phi
+ y_{\phi\chi_2}^{} \chi_2\chi_2\phi^*\!
+\text{h.c.} \),
\label{eq:FDM}
\end{eqnarray}
where the parameters
$(m_{\widehat{\chi}}^{},\,y_{\phi\chi_1}^{},\,y_{\phi\chi_2}^{}\!)$
are positive after proper phase rotations of
$(\phi,\,\chi_1,\,\chi_2)$.
As shown in Sec.\,\ref{sec:2.1}, the scalar field $\phi$ acquires a small VEV
$\,v_\phi^{}\!\simeq\! \lambda_{S\phi}^{}v_S^3/M_{\phi}^2$\,
due to the scalar seesaw from Eq.\eqref{eq:DL}.
The VEV $\,v_\phi^{}$ will induce additional Majorana masses for
$(\chi_1,\,\chi_2)$
through the Yukawa interactions in Eq.\eqref{eq:FDM}.
Thus, we have the following DM mass terms:
\beqa
\mathcal{L}_{\chi_1\chi_2}^{} \supset
- m_{\widehat{\chi}}^{}\chi_1\chi_2
+ \delta m_1^{} \chi_1\chi_1
+ \delta m_2^{} \chi_2\chi_2 +\text{h.c.},
\eeqa
where
$(\delta m_1,\,\delta m_2) \!=\!
(y_{\phi\chi_1}^{}\!\!v_\phi^{},\,y_{\phi\chi_2}^{}\!\!v_\phi^{})$.
To rotate $(\chi_1^{},\,\chi_2^{})$ into the mass-eigenstates
$(\chi^{},\,\chi')$, we make the following decomposition:
\begin{eqnarray}
\chi_1 \simeq
\fr{1}{\sqrt2}\(\chi^{}\!-\!\ii\chi'\),
\hspace*{4mm}
\chi_2 \simeq
\fr{1}{\sqrt2}\(\chi\!+\!\ii\chi'\).
\end{eqnarray}
In the limit $\,v_\phi^{}\!\ll\! m_{\widehat{\chi}}^{}$\,,
we derive the following Majorana masses for the
DM mass-eigenstates $(\chi,\,\chi')$:
\vspace*{-1mm}
\beqa
m_{\chi}^{}
&\,\simeq\,&
m_{\widehat{\chi}}^{} \!- ( \delta m_1\!+\!\delta m_2)\,,
\hspace*{5mm}
\nn
\\[-2mm]
\\[-2mm]
m_{\chi'}^{}
&\,\simeq\,&
m_{\widehat{\chi}}^{} \!+ ( \delta m_1\!+\!\delta m_2)\,,
\hspace*{5mm}
\nn
\eeqa
which have a mass-splitting,
\beqa
\Delta m_\chi^{} \simeq\,
2( \delta m_1\!+\!\delta m_2)\,.
\eeqa
To realize the required mass-splitting of $O({\rm keV})$ for explaining
the XENON1T anomaly, we choose the natural sample inputs,
$\lambda_{S\phi}^{},\, y_{\phi\chi_1}^{}\!,\,y_{\phi\chi_2}^{}
\!\!= O(0.01)$,
$\,v_S^{}\!\!=\!O(20)$ GeV, and
$M^{}_\phi\!\!=\!O(\text{TeV})$.
With these, we deduce a small VEV $\,v_\phi^{}\!=O(0.1)$MeV
and thus the desired mass-splitting
$\,\Delta m_\chi^{}\!=O(\text{keV})$.
From Eq.\eqref{eq:FDM}, we deduce the
$U(1)_X$ gauge interactions for the DM fields
$(\chi,\,\chi')$:
\beqa
{\cal L}_{\text{int}}^{} \,\supset\,
\ii q_{\widehat{\chi}}^{} g_X^{}
\!\(\chi^\dagger
\bar{\sigma}_\mu^{} \chi'\! -
{\chi'}^\dagger
\bar{\sigma}_\mu^{} \chi\)\!
{\cal X}^\mu .
\label{eq:FDM_vertex}
\eeqa
We note that the diagonal vertices
$\chi^{}$-$\chi^{}$-${\cal X}^\mu$ and
$\chi'$-$\chi'$-${\cal X}^\mu$ vanish,
whereas the above non-diagonal vertices can induce
the desired inelastic scattering for explaining
the XENON1T anomaly of DM-electron recoils.

\vspace*{1mm}

\begin{figure*}[t]   
\centering
\hspace*{-6mm}
\includegraphics[height=6.5cm,width=8.5cm]{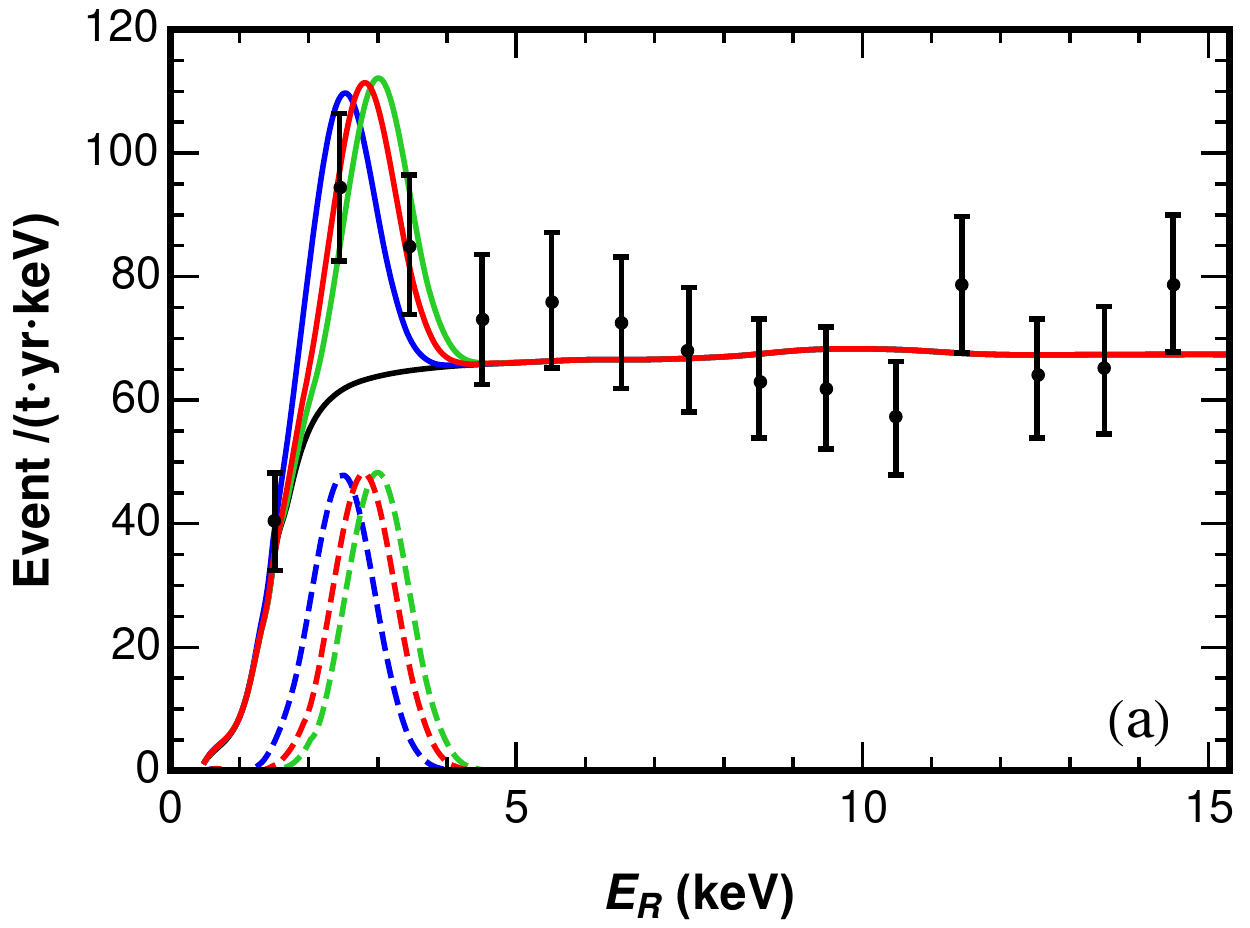}
\includegraphics[height=6.8cm,width=8.8cm]{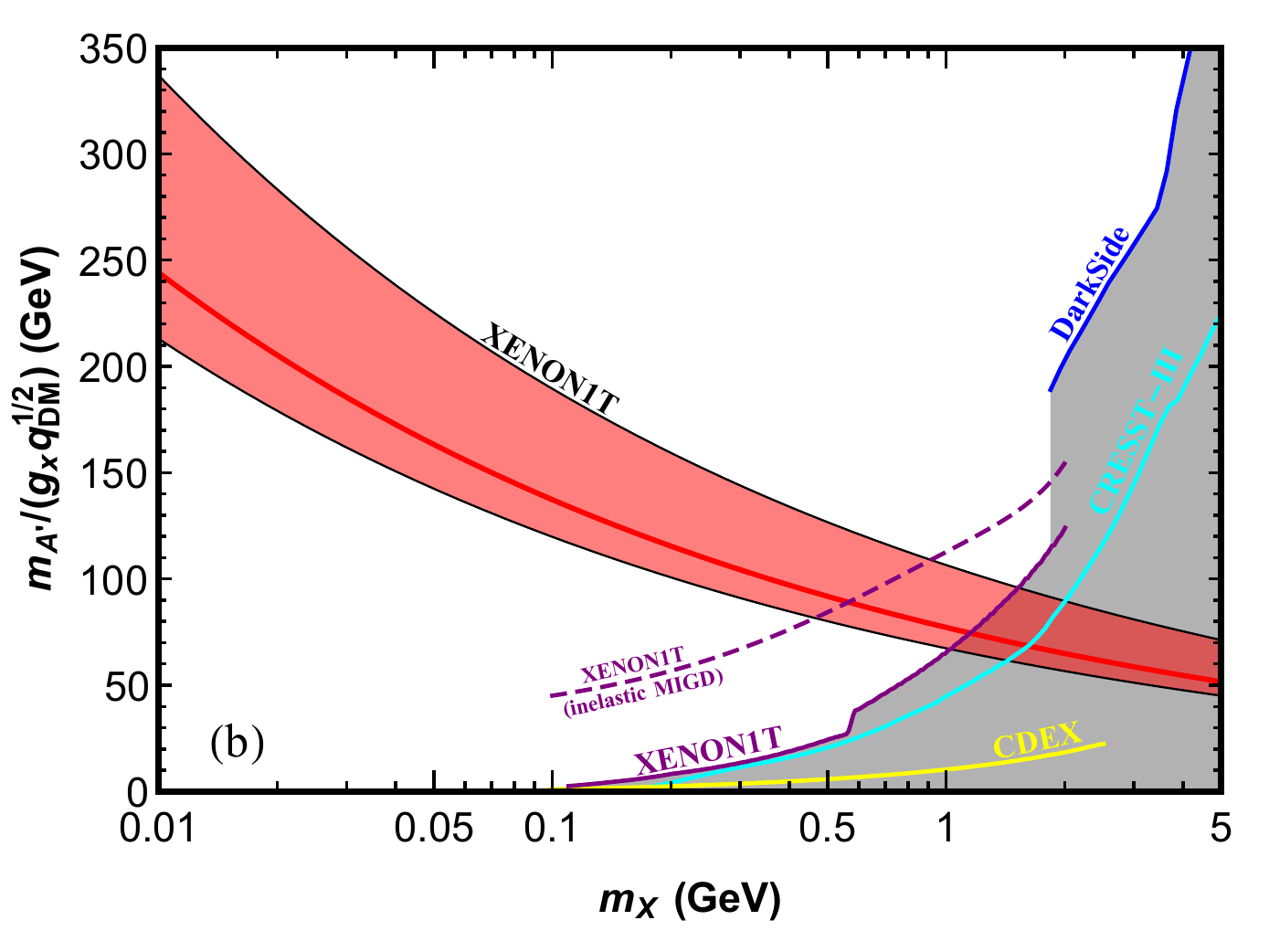}
\vspace*{-2.6mm}
\caption{\small
Plot\,(a):~Electron recoil energy spectrum from electron-DM scattering as predicted by the inelastic DM and
measured by XENON1T\,\cite{Aprile:2020tmw}. The (green, red, blue) dashed curves
present the inelastic DM contributions with
the DM mass-splitting $\,\Delta m_{\!X}\!=\!(2.5,\,2.8,\,3.0)$\,keV,
while the (green, red, blue) solid curves further include the XENON1T background
contribution (shown as the black solid curve).
The data points with error bars show the XENON1T measurement\,\cite{Aprile:2020tmw}.
Plot\,(b):~Constraints on our model by the direct DM detections of XENON1T, CRESST-III,
CDEX-1B and DarkSide. The pink area shows the allowed region ($95\%$\,C.L.)
and the red solid curve gives the central value
from fitting the XENON1T electron recoil data\,\cite{Aprile:2020tmw}.
The gray regions are excluded at $90\%$\,C.L.\,by
the nucleus recoil detections (of low threshold),
where the (purple,\,light-blue,\,green,\,dark-blue) curves
present the existing exclusion limits from
the XENON1T\,\cite{Aprile:2019jmx}, CRESST-III\,\cite{Abdelhameed:2019hmk},
CDEX-1B\,\cite{CDEX-1B}, and DarkSide-50\,\cite{DarkS} experiments
via nucleon recoils, respectively. 
The purple dashed curve presents 
the estimate of an improved constraint of Migdal effect 
on the inelastic DM inferred from the 
XENON1T result\,\cite{Aprile:2019jmx}.}
\label{fig:DD}
\label{fig:1}
\vspace*{3mm}
\end{figure*}

Most discussions in the rest of this paper can apply to both
the scalar and fermionic DM. For the simplicity of notations,
unless specified otherwise,
we will use the same symbols $(\Delta m_{\!X},\, \mX)$
and $(\widehat{X},\,X,\, X')$ for both scalar and fermionic DM.
For the fermionic DM, these notations refer to
$(\Delta m_\chi,\, m_{\widehat{\chi}}^{})$ and
$(\widehat{\chi},\,\chi,\,\chi')$.

\section{\hspace*{-2.5mm}Analyzing Constraints of DM Direct Detections}
\label{sec:DD_Event_Rates}
\label{sec:3}

For the present study, we derive the DM-electron inelastic scattering
cross section with zero-momentum-transfer, which holds
for both the scalar DM and fermionic DM:
\beqa
\bar\sigma_e^{} \equiv
\sigma_{Xe}^{}(|{\bf q}|\!=\!0) \,=\,
\frac{~q_e^2 q_{\text{DM}}^2 g_X^4\,}{4\pi}
\frac{m_e^2}{\,m_{A'}^4\,}\,,
\hspace*{10mm}
\label{eq:DD_sigma}
\eeqa
where $e_R^{}$ carries $U(1)_X$ charge
$\,q_e^{}\!\!=\!-\fr{1}{2}\,$.
The $U(1)_X$ charge of the scalar DM is
$\,q_{\text{DM}}^{}\!\!=\!q_{\widehat{X}}^{}\!=\!3$\,,
and the fermionic DM has
$\,q_{\text{DM}}^{}\!\!=\! q_{\widehat{\chi}}^{}\!\!=\!\!\fr{3}{2}$\,.
In Ref.\,\cite{He:2020wjs}, we proposed an effective field theory (EFT) approach
to perform a model-independent fit of the inelastic DM
for the XENON1T electron recoil spectrum\,\cite{Aprile:2020tmw}.
Our fit shows that the XENON1T anomaly\,\cite{Aprile:2020tmw}
can be fully explained by the two-component inelastic DM with mass-spitting
$\,\Delta m_{\!X}\!\equiv\! m_{X'}^{}\!-\!m_X^{}\!=\!2.8^{+0.2}_{-0.3}$\,keV
(68\%\,C.L.) and
$\,2.1\,\text{keV}\!\!<\!\! \Dm_{\!X}\!<\! 3.3\,\text{keV}$
(95\%\,C.L.).
We obtain the ratio of the inelastic cross section over the DM mass,
{ $\,\bar\sigma_e^{}/m_X^{}\!=\! (5.7\pm2.6)$ $\!\times 10^{-44}{\rm cm^2/GeV}$,}
under the condition that the DM density
contains equal amount of $X$ and $X'$ particles.
Thus, by fitting the XENON1T data\,\cite{Aprile:2020tmw},
we derive the following bound on our model:
\vspace*{-1.5mm}
\beqa
\hspace*{-6mm}
\frac{\,m_{A'}^{}}{g_X^{}} = 1.3^{+0.3}_{-0.1}
\!\times\! 10^2
 \,\text{GeV}\!\times\!
\left[\left|
\frac{q_e q_{\text{DM}}}{3/2}
\right|^{\frac{1}{2}}\!\!
\(\!\frac{1{\rm GeV}}{m_X}\!\)^{\!\!\frac{1}{4}}\right]\!.~
\label{eq:XENON1T_vs}
\eeqa

In Fig.\,\ref{fig:1}(a), we present the electron recoil energy spectrum
as predicted by the inelastic DM, in comparison with that of
the XENON1T measurement\,\cite{Aprile:2020tmw}.
The recoil spectra plotted in the (green,\,red,\,blue) dashed curves
correspond to the inelastic DM contributions with
mass-splitting $\,\Delta m_{\!X}\!=\!(2.5,\,2.8,\,3.0)$\,keV,
whereas the (green, red, blue) solid curves further include the  background contribution as given by model $B_0^{}$ of XENON1T\,\cite{Aprile:2020tmw}
 (depicted by the black solid curve).
The data points with error bars show the XENON1T measurement\,\cite{Aprile:2020tmw}.
We have input a sample cross-section/mass ratio 
$\,\over{\sigma}_e^{}/\mX\! =5.7 \!\times\! 10^{-44}\text{cm}^2/$GeV,
which is the best fit with XENON1T data\,\cite{Aprile:2020tmw}
as mentioned above Eq.\eqref{eq:XENON1T_vs}.
We note that the shape of our fitted recoil peak is mainly determined by
the data points below 5\,keV and the fit with the spectrum above 5\,keV
has little effect on the shape of this peak.
The background model $B_0^{}$ is fitted to the recoil energy data over a very wide range of $(1\!-\!210)$\,keV\,\cite{Aprile:2020tmw}, hence its 
normalization has negligible fluctuation.

\vspace*{1mm}

In Fig.\,\ref{fig:1}(b), we show
$\,m_{\!A'}^{}/(g_X^{}\!\sqrt{q_{\text{DM}}^{}})$\,
as a function of the DM mass $m_{\!X}^{}$\,.
By fitting the recent XENON1T data\,\cite{Aprile:2020tmw}, we present
the allowed parameter space by the pink area at $95\%$\,C.L.,
and plot the central values (best fit) by the red solid curve.
PandaX-II collaboration has accumulated 100.7 ton-day data from
measuring the electron recoil spectrum\,\cite{PandaX2}
which is consistent with the background fluctuations.
In each bin of the PandaX-II data, we find the expected DM signal rate
to be less than 10 events/keV, which is smaller than the error bar.
Hence, the best fit of our model is consistent with PandaX-II
although its current bound is too weak to be shown
in the same Fig.\,\ref{fig:1}.

\vspace*{1mm}

The DM particles in our model can also scatter with nuclei by exchanging
the mediator $A'_\mu$ since the quarks $(u_R^{},d_R^{})$ carry $U(1)_X$
charges as in Table\,\ref{tab:1}.
The DM-nucleon scattering cross section is dominated by the vector coupling
of $A'_\mu$ with quarks since the contribution of the axial-vector coupling 
is suppressed by the velocity of the recoiled nucleus. 
Thus, the spin-independent contribution dominates the DM-nucleon scattering. 
The $U(1)_X$ charge of a nucleon is the sum of its valence quarks 
at low energy, i.e., 
$q_p^{}\!=\!\fr{1}{2}$\, for proton and 
$q_n^{}\!=\!-\fr{1}{2}$\, for neutron. 
We thus compute the DM-nucleon scattering cross section
as follows\,\cite{Belanger:2008sj,Yu:2011by},
\vspace*{-1mm}
\beqa
\sigma_{\!X\!N}^{} \,=\,
\frac{m_N^2 m_X^2 q_N^2 q_{\text{DM}}^2{g_X^4}}
{\,4\pi(m_X^{}\!+\!m_N^{})^2 m_{A'}^4\,}\,,
\label{eq:DD_X-N}
\eeqa
where $\,m_N^{}\!\simeq 940$\,MeV is the mass of a nucleon, 
and $\,q_N^{}\!=\!\fr{1}{2}\,(-\fr{1}{2})$ for protons (neutrons).
The above formula of the DM-nucleon
scattering cross section remains the same 
for both scalar and fermionic DM,
except that the $U(1)_X$ charge $q_\text{DM}^{}$ of the DM 
can differ in the two cases.
This is consistent with the fact that the dominant contribution to the scattering is spin-independent and does not rely on the mass-splitting $\Delta \mX$.
If such scattering could flip the spin of the fermionic DM,
it must proceed by changing either the orbital angular momentum of the system
or the spin state of the nucleus. The amplitude of the former is suppressed
by the low DM velocity, while the latter depends on the nuclear spin and
thus is much smaller than the dominant spin-independent contribution.
The leading amplitude is thus diagonal in the space of DM spin-states,
and is proportional to
$J_X^\mu \!=\! \langle X(k_2^{})|\hat{J}^\mu_X |X'(k_1^{})\rangle
\!\propto q_\text{DM}^{}(k^\mu_1 \!+ k^\mu_2) + {\cal}O(\Delta m_{\!X})$.
This form is fixed by the $U(1)_X$ current conservation
$\,q_\mu^{} J_X^\mu\!=\!0\,$ in the limit
$\,\Delta\mX\!\!\rightsquigarrow \!0\,$,
where the momentum transfer
$\,q^\mu$ $\!\!=\!k^\mu_1 \!\!-\! k^\mu_2$.\,
This form does not depend on the spin of DM,
hence the leading term of the DM-nucleus scattering cross section
remains the same for both the scalar and fermionic DM.

\vspace*{1mm} 

We also note that the $U(1)_X$ charges of protons and neutrons 
have opposite signs in our model. 
This gives the total $U(1)_X$ charge of a nucleus $\,Q\!=\!\frac{1}{2}(A\!-\!2Z)$\,,
where $A$ and $Z$ denote the nucleon number and proton number, respectively. 
The form factor is nearly constant,\ 
$f(qr_\text{n}^{})\!\simeq\!1$\,, 
because for the light DM $\,\mX\!<\!5\text{GeV}$\, 
the momentum transfer 
$\,q\!\lesssim\! O(10\text{MeV})$\, is small. 
Thus, to compare our model with the nuclear recoil measurements 
which usually assume isospin symmetry between protons and neutrons for spin-independent interactions,
we can properly rescale their original bounds on the DM-nucleon cross section by a factor of $\,A^2/(A\!-\!2Z)^2$,\,
where a weighted average over isotope abundance is understood. 

\vspace*{1mm}

For dual-phase detectors
such as XENON1T \cite{Aprile:2018dbl}
and DarkSide-50\,\cite{DarkS-2015},
the nuclear recoil causes primary scintillation signals
and are measured with energy thresholds at several keV,
which are sensitive to DM particles with masses $\gtrsim\!5\,$GeV.
Due to Migdal effect\,\cite{Migdal},
the recoiled nuclei produce ionization and/or excitation of
their atomic electrons with finite probability.
The secondary radiation signals created by these electrons
can be detected with much lower energy thresholds ($\lesssim\! 1\,$keV),
and thus largely enhance the sensitivity to sub-GeV DM. 
These effects have been included in the direct detection  experimental papers such as\,\cite{Aprile:2019jmx}. 
The differential rate of electron recoil (ER) in the Migdal process accompanying an nucleus recoil (NR) is given by\,\cite{Migdal}\cite{Aprile:2019jmx},
\begin{eqnarray}
\label{eq:dR/dE-Migdal}
\hspace*{-8mm}
\frac{{\rm d}R}{{\rm d}E_{\rm ER}}\! \simeq \! 
\int\!\! {\rm d}E_{\rm NR} {\rm d}v\!
\left[
\frac{{\rm d}^2 R_{\rm N}}{\,{\rm d}E_{\rm NR} {\rm d}v\,}
\sum_{n,\ell} 
\frac{{\rm d}P^c_{q_e}}
{\,2\pi\, {\rm{d}}E_{\rm ER}\,}  
\right]\!,
\end{eqnarray} 
where $P^c_{q_e}$ encodes the probability for the ionization of atomic electron with quantum numbers $(n,\ell)$ suppressed in the expression.  The differential rate of the nuclear recoil,
$\frac{\d^2 R_{\rm N}}{\,\d E_{\rm NR} \d v\,}$,\,
depends only on the short scale particle physics model 
following the standard calculation of DM-nuclear scattering. 
Because of the clear separation of the nuclear and atomic physics
effects as in Eq.\eqref{eq:dR/dE-Migdal}, 
the XENON1T collaboration already took into account the atomic ionization factor $P^c_{q_e}$ and transferred its limit on 
the electron recoil event rate to the limit on the DM-nucleon 
elastic cross section. 
For exothermic inelastic scattering, the energy release from the DM  makes it easier to excite Migdal electrons 
than elastic scattering as shown in \cite{Bell:2021zkr}. 
Thus, we can view the XENON1T result\,\cite{Aprile:2019jmx} on DM-nucleus elastic scattering as a conservative constraint 
on Eq.\eqref{eq:DD_X-N}.
So far in the literature the constraints 
of including Migdal effect are 
derived only for elastic DM\,\cite{Aprile:2019jmx} and for inelastic DM with a few special inputs of $\delta$\,\cite{Bell:2021zkr}.
In Appendix\,\ref{sec:APD:inelastic_migdal_est}, we make an estimate of the improved constraint of Migdal effect on the inelastic DM 
by using the XENON1T result\,\cite{Aprile:2019jmx}.

\vspace*{0mm}

In the following, we summarize the best constraints on DM-nucleon  scattering from current direct detection experiments.
The strongest bound is set by
the recent XENON1T detection\,\cite{Aprile:2019jmx},
which probes the light DM with masses down to about 85\,MeV
by measuring electronic recoils induced by the Migdal effect and bremsstrahlung.
It can detect both scintillation and ionization signals as well as
ionization signals only which allows 
for a lower detection threshold.%
\footnote{ 
We thank our experimental colleagues Qing Lin and
Jingqiang Ye in the XENON1T Collaboration\,\cite{Aprile:2019jmx}
for discussing their analysis of including Migdal effect
and confirming our application of their bound from Fig.5 of 
Ref.\,\cite{Aprile:2019jmx}.\  We also thank Qian Yue of the
CDEX-1B Collaboration for confirming our application of 
their experimental bound\,\cite{CDEX-1B} 
including Migdal effect.}\
The recent DarkSide-50 measurement\,\cite{DarkS}
used a target of low-radioactivity argon and analyzed the ionization signals,
which probes the light DM mass down to $(1.8\!-\!6)$\,GeV range.
The CDEX-1B experiment\,\cite{CDEX-1B} uses $P$-type point contact
germanium (PPCGe) detectors to detect both the nuclear recoil energy
and electron ionization energy.
It can probe the light DM mass down to 50\,MeV.
Utilizing solid-state detectors with low energy thresholds
is another significant means for light DM detection.
The CRESST-III detection\,\cite{Abdelhameed:2019hmk} operates scintillating
$\rm CaWO_4$ crystals as cryogenic calorimeters
and can achieve a low nuclear recoil threshold energy of 30.1\,eV.\,
It is sensitive to the light DM of mass below $\sim\!\!2$\,GeV.\,

In Fig.\,\ref{fig:1}(b), we present the bounds (90\%\,C.L.) from
XENON1T\,\cite{Aprile:2019jmx}, CRESST-III\,\cite{Abdelhameed:2019hmk},
CDEX-1B\,\cite{CDEX-1B}, and DarkSide-50\,\cite{DarkS}
on the parameter space of our model, shown by the gray region,
over the mass range $\,m_X^{}\!=(0.01\!-5)$\,GeV.\,
Here the purple, light-blue, yellow, and dark-blue curves give the
90\% exclusion limits set by
the XENON1T, CRESST-III, CDEX-1B, and DarkSide-50 experiments, respectively.
We find that the XENON1T measurement (including Migdal effect) 
imposes the strongest limit on our parameter space 
among the existing bounds.\footnote{%
Note that for DM mass $\lesssim\!\! 1$\,GeV, 
the typical recoil energy of scattering off
an nucleus is below keV. We thus only include the
down scattering process $X'\!+\!N\!\to\! X\!+\!N$\, for the bound.}
This constrains the inelastic DM mass within the range
{$\,m_X^{}\!\lesssim\! 1.5$\,GeV.}
The purple dashed curve shows our estimated constraint 
of Migdal effect on the 
(exothermic) inelastic DM inferred from the original XENON1T result\,\cite{Aprile:2019jmx} on elastic DM based on the scaling relation of Eq.\eqref{eq:Migdal_scaling}. We see that this 
improved analysis could enhance the constraint to around  
{$\,\mX\!\lesssim\! 0.9$\,GeV}. We note that in general using the 
Migdal effect can place strong constraints on inelastic DM 
and our new method can be applied to a broad class of inelastic DM models. We will perform a systematic analysis of Migdal effect on various inelastic DM models in the future work.

\section{\hspace*{-2.5mm}Analyzing Cosmological Constraints}
\label{sec:CosmoConstraints}
\label{sec:4}

In this section, we analyze the relevant cosmological constraints on
the present inelastic DM model.

\vspace*{-2mm}
\subsection{\hspace*{-2.5mm}Lifetime of the Heavier DM Component 
\boldmath{$X'$}}
\label{sec:SDM}
\label{sec:4.1}
\vspace*{-2mm}

To resolve the XENON1T anomaly by inelastic DM, it is important to
ensure that the lifetime of the heavier DM component $X'$ is longer than
the age of the Universe.
Since we have the DM mass-splitting
$\,\Delta m_{\!X} \!\ll\! m_e^{}$\,,
only the decays such as $\,X'\!\to\!X\gamma\gamma$\, and
$\,X'\!\to\!X\nu\bar\nu$\, are kinematically allowed.
The decay channel $\,X'\!\to\!X\gamma\gamma$\,
arises from one-loop diagrams with electron $e_R^{}$
or quarks $u_R^{}$\,($d_R^{}$) running in the loop.
Thus, with \cite{He:2020wjs,Jackson:2013pjq},
we can estimate the partial decay width of $X'$ as
\beqa
\hspace*{-8mm}
\Gamma_{X'\to X\gamma\gamma}^{}  \,\simeq\,
\!\!\sum_{f=e,u,d\,}\!\!\!
\frac{~\alpha^{2}g_X^4q_{\text{DM}}^2\,}{\,155520\pi^5\,}\!\!
\left(\!\!\frac{\,q_f^{} N_{cf}^{}\,}{m_f^2}\!\!\right)^{\!\!\!2}
\!\!
\frac{\,\Delta m_{\!X}^9}{~m_{A'}^4\,}\,,
\label{eq:X'-Xgaga}
\eeqa
where the sum runs over the electron-loop ($f\!=\!e$) and quark-loops
($f\!=\!u,d$), $q_f^{}$ denotes the electric charge of each fermion $f$,
and the color factor $N_{cf}^{}\!=\!1\,(3)$ for the electron (quark) loop.
Since the quark masses $\,m_{u,d}^{}\!\approx (1.5\!-6)$MeV
are much larger than the electron mass with the mass ratio
$\,(m_e^{}/m_{u,d})^4\!=\!O(10^{-2}\!-\!10^{-4})$,
we see that the electron loop
dominates the partial decay width \eqref{eq:X'-Xgaga}.
From fitting the XENON1T event rate, we have a small DM mass-splitting
$\,\Delta m_{\!X}\!\simeq\! 3$\,keV and the ratio of $\,m_{A'}^{}/g_X^{}$
given by Eq.\eqref{eq:XENON1T_vs}.
This leads to a tiny partial decay width
$\,\Gamma_{\!X'\to X\gamma\gamma}^{}\!\!\approx\!
(5\!\times\! 10^{24}\text{yr})^{-1}$,\,
and thus
$\,\Gamma_{\!X'\to X\gamma\gamma}^{-1}\!\ggg\! \tau_U^{}$,
where $\,\tau_U^{}\!\simeq\! 1.38\!\times\! 10^{10}$yr\,
is the age of the present Universe. Such a long lifetime is also much beyond the X-ray\,\cite{xray_DDM} and CMB\,\cite{CMB_DDM} constraints on the lifetime of radiatively decaying DM. 
These observations typically exclude the DM with lifetime 
shorter than $10^{20}\,$yr.

\vspace*{1mm}

The other decay channel $\,X'\!\!\to\! X\nu\bar\nu\,$
occurs through the $A'$ exchange due to the neutral gauge boson
mixing matrix \eqref{eq:boson_mixing}.
The contribution of the $Z$ exchange as induced by the mixing matrix
\eqref{eq:boson_mixing} is fully negligible because of the large suppression
factor $\,(m_{A'}^4/M_Z^4)\!\lesssim\! O(10^{-4})$\,
relative to the $A'$ exchange.
Thus, we can estimate the $X'$ decay width as follows:
\beqa
\Gamma_{X'\rightarrow X\nu\bar\nu}^{}  \,\simeq\,
\frac{~q_{\text{DM}}^2 g_X^4\,}{\,160\pi^{3}\,}
\frac{\,v_1^4\Delta m_{\!X}^5\,}{\,v_h^4m_{A'}^4\,}\,,
\eeqa
where we have included the contributions of the final state neutrinos
from three families. The result holds for both scalar\,(fermionic) DM.
By requiring the $X'$ lifetime be much longer than the age of
our present universe,
$\,\Gamma_{X'\to X\nu\bar\nu}^{-1}\!\gtrsim\! 10\tau_U^{} \!\!\gg\! \tau_U^{}$,\,
we derive a constraint:
\vspace*{-1mm}
{\beqa
v_1^{} \,\lesssim\, 19\,\text{GeV} \!\times\!
\(\!\!\frac{~1{\rm GeV}\,}{m_X^{}}\!\)^{\!\!\frac{1}{4}}.
\label{eq:X'stability}
\eeqa}
This shows that in our model the electroweak symmetry breaking (EWSB)
is mainly generated by the Higgs doublet $H_2^{}$, and the VEV ratio $\,\tan\beta=v_2^{}/v_1^{}\!\gtrsim 10\,$.

\begin{figure*}[t]
	\centering
	\includegraphics[height=7.5cm,width=8.9cm]{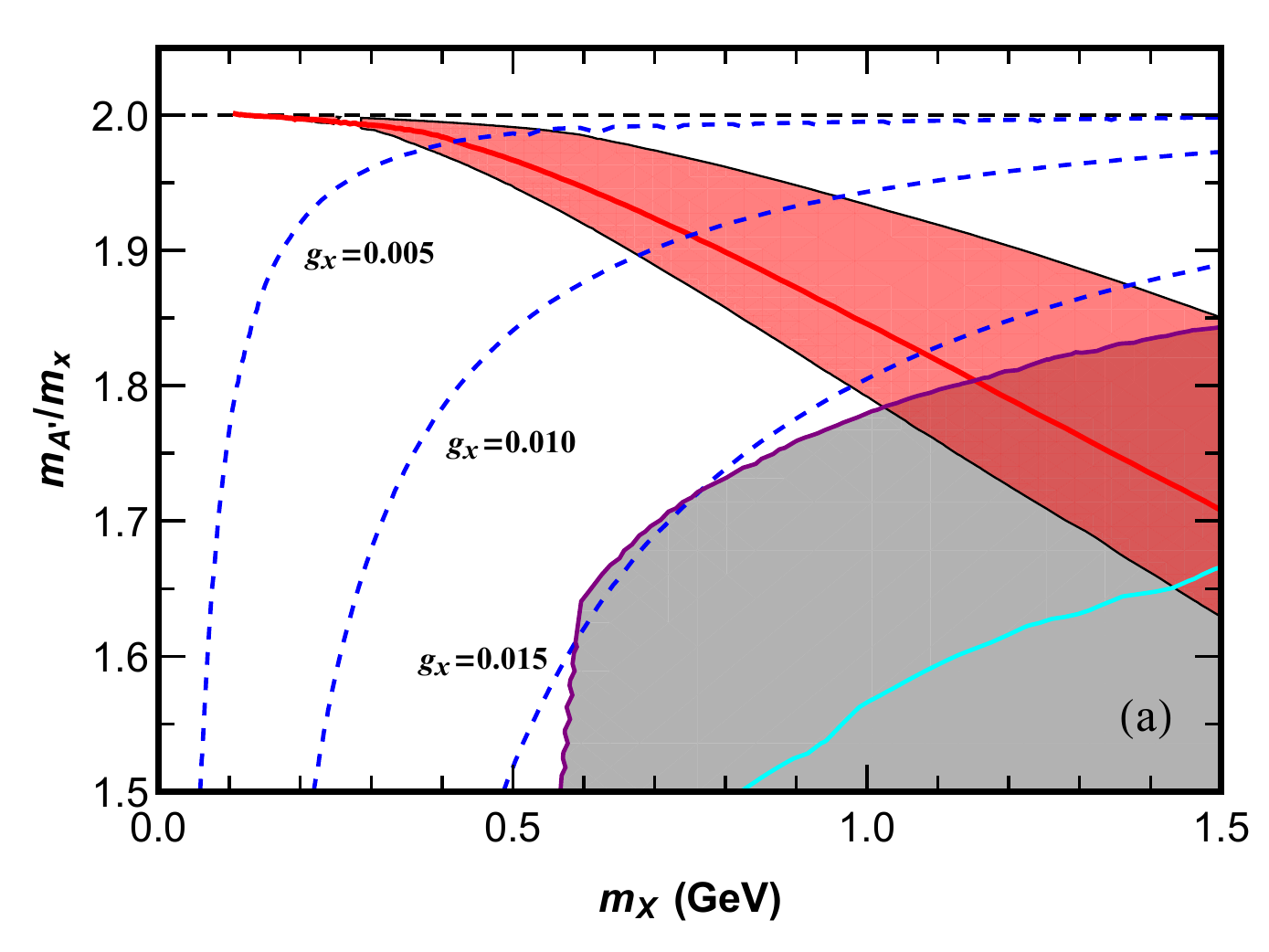}
	\hspace*{-3mm}
	\includegraphics[height=7.5cm,width=8.9cm]{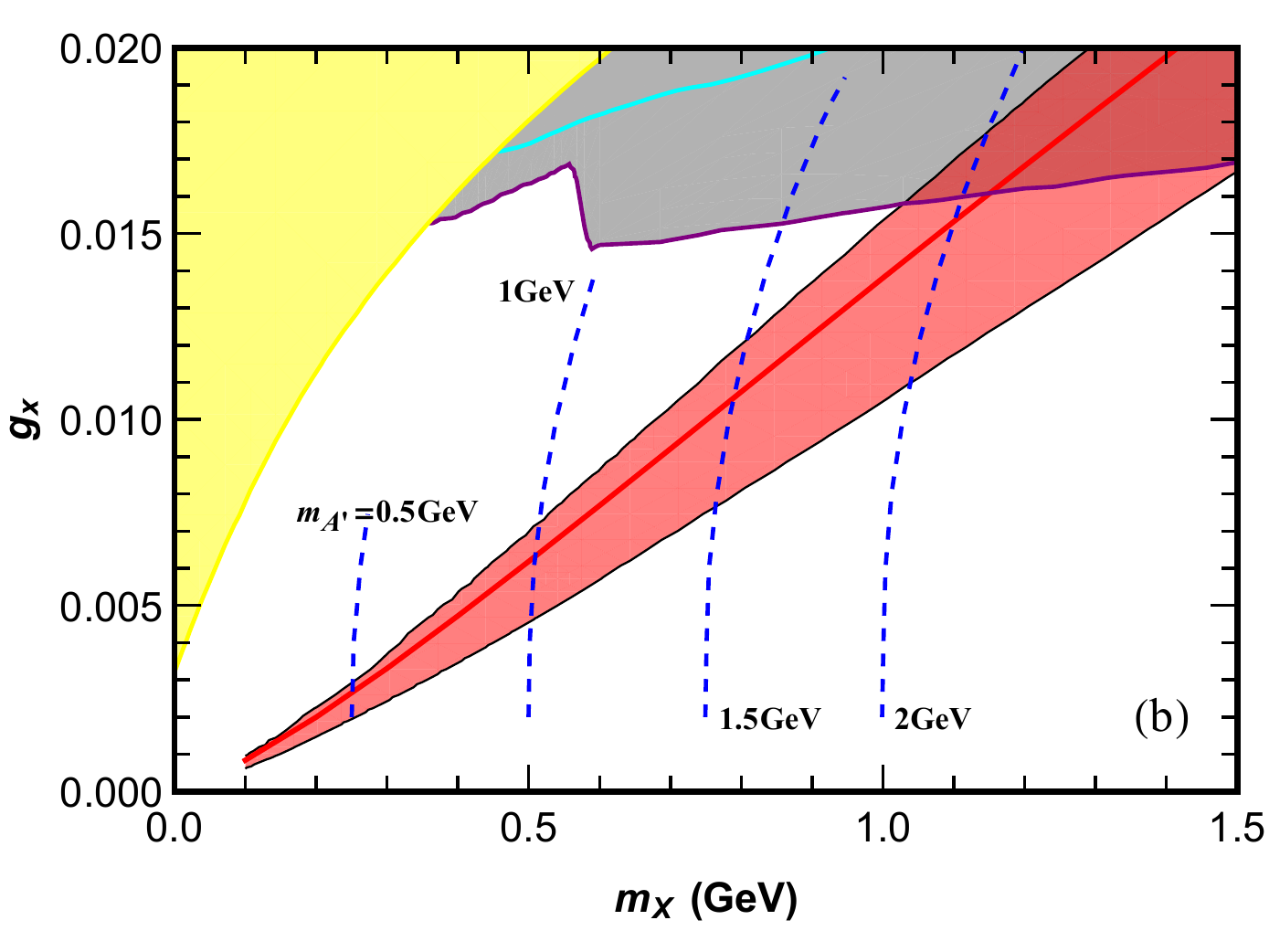}
	\vspace*{-2mm}
	\caption{\small{
			Allowed parameter space for the inelastic scalar DM.
			In each plot, the pink areas depict the allowed regions (95\%\,C.L.)
			by the XENON1T electron recoil data\,\cite{Aprile:2020tmw}
			combined with the DM relic density bound,
			and the red solid curve gives the best-fit values.
			The gray areas are excluded by nucleon recoil detections of low threshold
			(90\%\,C.L.),
			where the purple and light blue curves are exclusion limits
			by XENON1T\,\cite{Aprile:2019jmx} and CRESST-III\,\cite{Abdelhameed:2019hmk}, respectively.
			Plot\,(a):
			Constraints on the parameter space of $\,m_X^{}$ versus $m_{A'}/m_X^{}$\,.
			The blue dashed curves show the allowed parameter space of realizing the observed
			DM relic density for a set of sample $g_X^{}$ values.
			Plot\,(b):
			Constraints on the parameter space of $\,m_X^{}$ versus $g_X^{}$\,.
			The blue dashed curves show the parameter space of realizing the observed
			DM relic density for a set of sample mediator masses $m_{A'}^{}$.
			The yellow region is excluded by the DM relic density bound for
			$\,m_{A'}^{}/m_X^{}\!<2$\,.}
	}
	\label{fig:SRelic}
	\label{fig:2}
\end{figure*}

\noindent

\vspace*{-2mm}
\subsection{\hspace*{-2.5mm}Dark Matter Relic Abundance} 
\label{sec:SDM}
\label{sec:4.2}
\vspace*{-2mm}

Since our construction always holds
$\,\Delta m_{\!X} \!\ll\! m_X^{}$\,,
the two DM components can be regarded as a degenerate complex scalar
or Dirac spinor before they freeze out.
We thus determine the relic density of a complex scalar $\widehat{X}$
or a Dirac spinor $\widehat{\chi}$\,.
We compute the DM relic density of the current model 
by using the package
{\tt MicrOMEGAs}\,\cite{Belanger:2008sj,Belanger:2020gnr}
and present our results for the scalar DM model in Fig.\,\ref{fig:2}.

\vspace*{1mm}

In plot\,(a) of Fig.\,\ref{fig:2}, 
the blue curves show the masses of $X\,(X')$ and $A'$
which achieve the observed DM relic abundance,
for each given value of the gauge coupling $g_X^{}$.
The pink area presents the allowed regions (95\%\,C.L.)
by the XENON1T data\,\cite{Aprile:2020tmw}
combined with the DM relic density bound,
and the red solid curve depicts the best-fit.	
The gray region is excluded by the
XENON1T\,\cite{Aprile:2019jmx} and CRESST-III\,\cite{Abdelhameed:2019hmk}
measurements combined with the DM relic density bound.
We find that the DM mass range
$\,\mX\!\!\lesssim\!1.5$\,GeV is favored in our model,
and in the viable parameter space
the dominant DM annihilation channel is
$\,\widehat{X}^*\widehat{X}\!\!\to\!\!{A'}\!\!\!\to\! f_R^{}\bar{f}_R^{}$\,,
where $f\!=\!u,d,e$\,.
In plot\,(b), the pink region presents the allowed 
parameter space of $\,m_{X}^{}$ versus $\,g_X^{}$ at 95\%\,C.L., 
and the red solid curve gives the best-fit values.
The blue curves show the allowed values of the gauge coupling $\,g_X^{}$ as a function of the DM mass $\,m_X^{}$,
which achieve the observed DM relic abundance
for each given mediator mass $m_{A'}^{}$.
We fit the XENON1T electron recoil data\,\cite{Aprile:2020tmw}
and DM relic density bound together,
and include the constraints of
{$\,m_X^{}\!\!\lesssim\! 1.5$\,GeV}
by XENON1T\,\cite{Aprile:2019jmx} and
CRESST-III\,\cite{Abdelhameed:2019hmk} nuclear recoil measurements
(the gray area).
With these, we derive the combined limit
{$\,g_X^{}\!\!\lesssim 0.017$\,}
for the case $\,m_{A'}^{}/m_X^{}\!\!<\!2$\,.

\vspace*{1mm}

In Fig.\,\ref{fig:3}, we present our results for the inelastic fermionic DM.
Similar to the scalar DM case, the annihilation of fermionic DM
in the viable parameter space is dominated by the $s$-channel process
$\,\widehat{\chi}^\dagger \widehat{\chi} \!\!\to\!\!
{A'}\!\!\!\to\!\! f_R^{}\bar{f}_R^{}$\,.
But, the annihilation is more efficient in the case of
fermionic DM because it is $s$-wave dominant;
whereas for the scalar DM, the process is $p$-wave dominant due to
its derivative coupling.
In order to achieve the observed DM relic density,
the parameters for the fermionic DM have to be farther away from
the resonance region than that for the scalar DM.
Hence, in Fig.\,\ref{fig:3}(a) the viable parameter region (pink area)
for the fermionic DM covers lower values of the mass ratio
$m_{A'}^{}/m_\chi^{}$,
as compared to that in Fig.\,\ref{fig:2}(a).

\vspace*{-4mm}
\subsection{\hspace*{-2.5mm}Current Composition of 
Inelastic Dark Matter}
\label{sec:SDM}
\label{sec:4.3}
\vspace*{-2mm}

%
\begin{figure*}
	\centering
	\includegraphics[height=7.5cm,width=8.9cm]{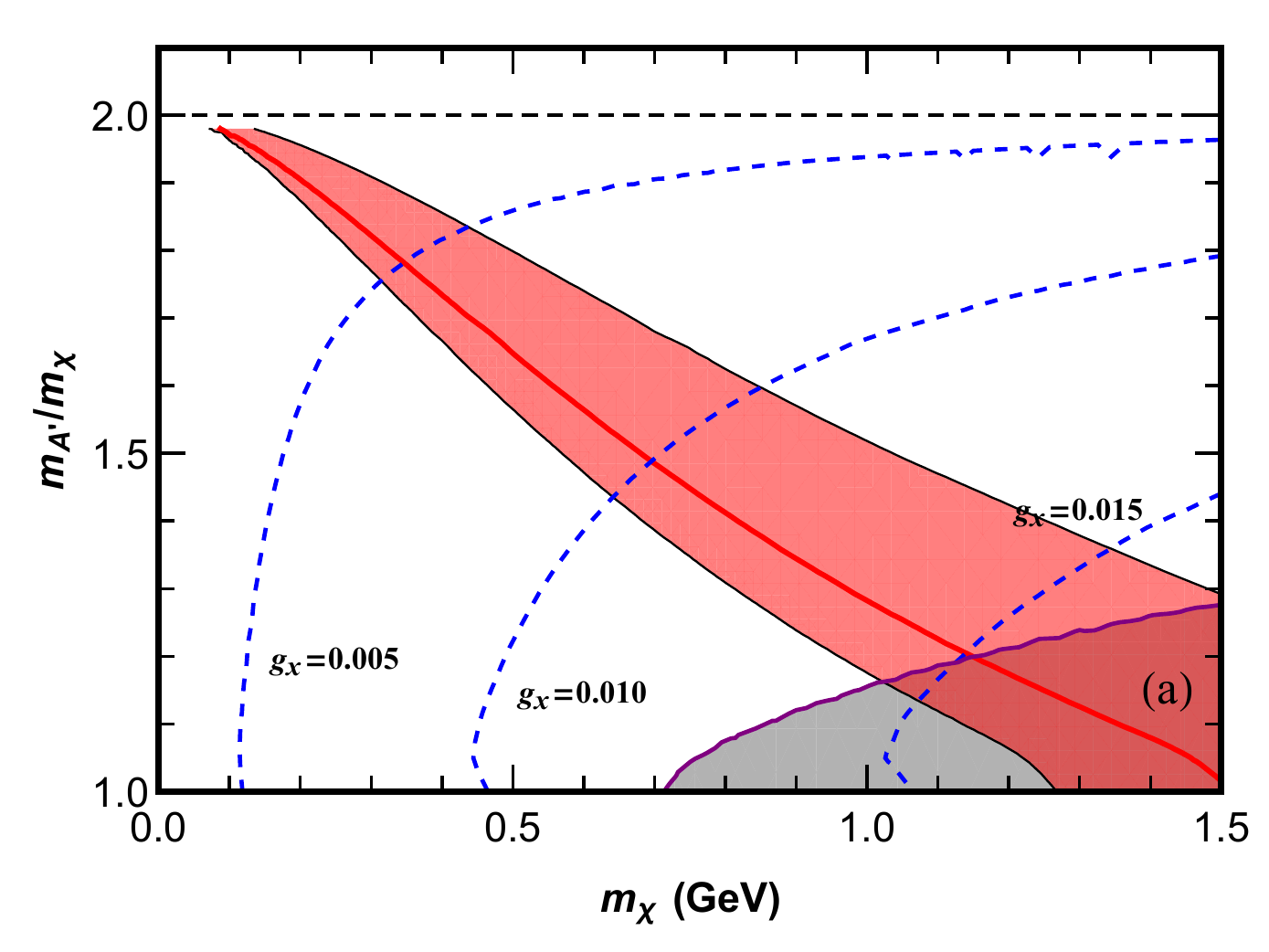}
	\hspace*{-2mm}
	\includegraphics[height=7.5cm,width=8.9cm]{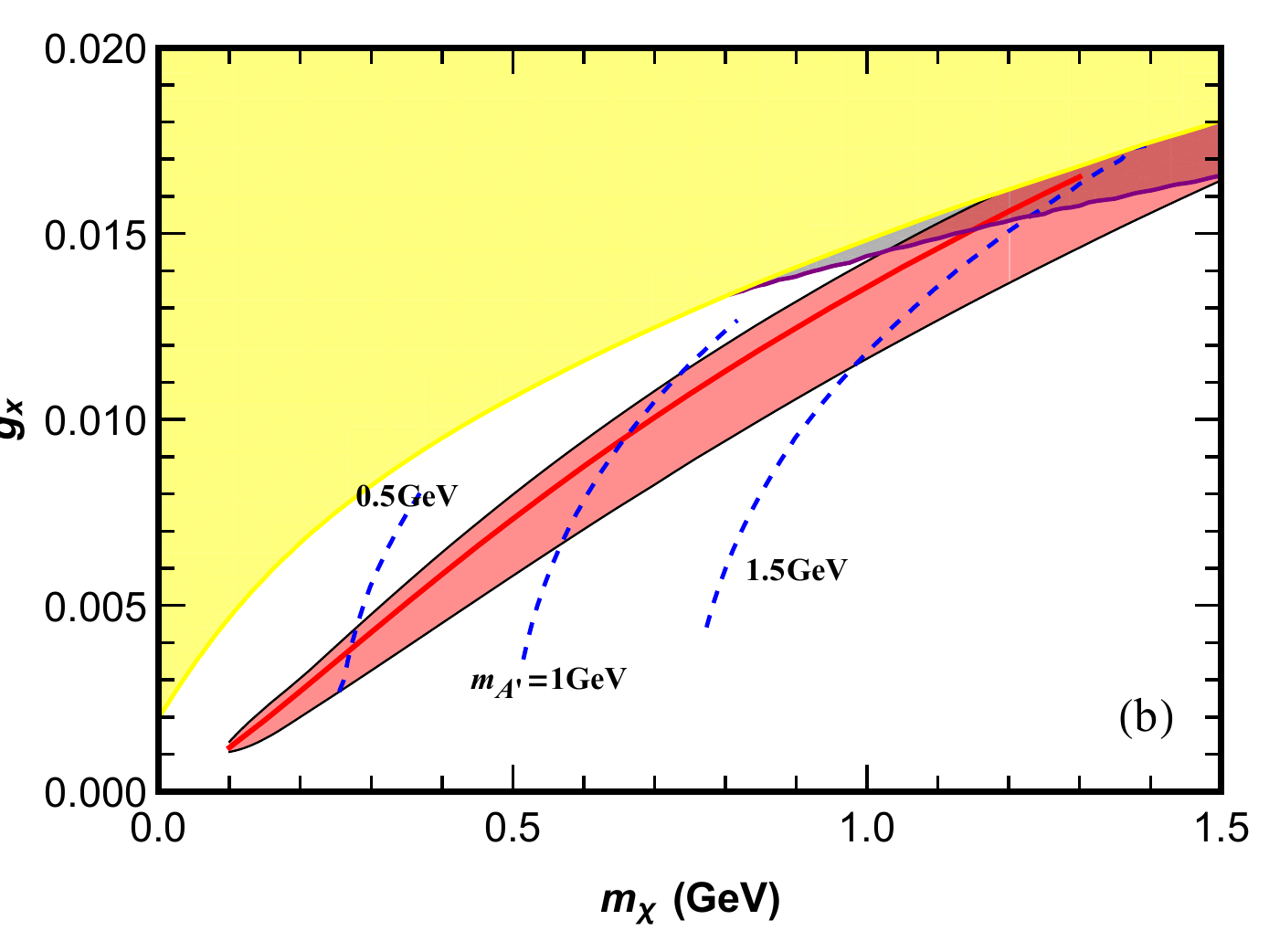}
	\vspace*{-2mm}
	\caption{\small
		Allowed viable parameter space for the inelastic fermionic DM,
		as shown in plots\,(a) for $\mX$ versus $\mAP /\mX$, and
		in plots\,(b) for $\mX$ versus $g_X^{}$.
		The curves and shaded regions have the same meanings as defined
		in the caption of Fig.\,\ref{fig:2} except the current plots present
		the case of the inelastic fermionic DM.
	}
	\label{fig:FRelic}
	\label{fig:3}
	\vspace*{-2mm}
\end{figure*}

Next, we examine whether the constrained couplings
in Figs.\,\ref{fig:2}-\ref{fig:3}
are consistent with the condition $\,n_X^{}\!=\!n_{X'}^{}$\,.
After the decoupling of the annihilation process
$\,\widehat{X}\widehat{X}^\dagger \!\!\to\!\! f_R^{}\bar{f}_R^{}$,\,
the total DM number density
$\,n_{\widehat{X}}^{}\!=$ $n_X^{}\!+\!n_{X'}^{}$\,
is fixed.
But, $X$ and $X'$ can still convert into each other
via processes $\,e^\pm X \!\leftrightarrow\! e^\pm X'$\,
and
$\,X'X'\!\!\leftrightarrow\!\! X X$\,.
As we showed previously\,\cite{He:2020wjs}, the former process decouples
at $\,T\!\sim\! m_e^{} \,(\gg\! \Delta m_{\!X})\,$
for the GeV scale DM.
The latter process is induced by $t$- and $u$-channel exchanges of $A'$.
If the annihilation $\,X'X'\!\!\to\!\! XX\,$
decouples at a temperature $\,T'\!\!<\!\!\Delta m_{\!X}$\,,
the $X'$ density would be exponentially suppressed
by a factor of $\,\exp(-\Delta m_{\!X}/T')$\,.
As an estimate, this process becomes inefficient
when the reaction rate $\,\Gamma(T')\!\lesssim\! H(T')$\,.
We estimate this reaction rate as
$\,\Gamma(T')\!\simeq\! \langle\sigma' v_{\text{DM}}^{}\rangle\,
n_{\text{DM}}^{}$,\,
in which the thermally averaged DM annihilation cross section
$\,\langle\sigma' v_{\text{DM}}^{}\rangle \!\simeq g_X^4 q_{\text{DM}}^4
\!\sqrt{m_X^3 T'\,}\! /(\pi m_{A'}^{4})$\,
and the DM density
$\,n_{\text{DM}}^{}\!\simeq\!T_{\text{eq}}^{}{T'}^3\!/m_X^{}$\,.
Here $T_\text{eq}^{}$ is the temperature at the matter-radiation equality.
Thus, with these and Eq.\eqref{eq:XENON1T_vs}, we can estimate
\beqa
T' \,\approx\, 0.014\,\text{GeV} \!\times\!
\(\!\! \frac{\,1\hspace*{0.2mm}\text{GeV}\,}{m_X}\!\)
\!\times\!
\(\!\! \frac{\,3\,}{\,q_\text{DM}^{}\,}\!\!\)^{\!\!\frac{4}{3}}\!.
\hspace*{5mm}
\eeqa
Shortly after the kinetic decoupling, the DM temperature
drops rapidly as $a(t)^{-2}$ and falls below $\Delta m_{\!X}$ quickly.
The $X'$ density would get depleted if the annihilation
$\,X'X'\!\!\to\! XX\,$ is still efficient.
As a conservative estimate, we demand $\,T'\!\gtrsim\! m_e^{}$,
so the process $\,X'X'\to\! XX\,$
decouples before the kinetic decoupling between $e^\pm$ and the DM.
This imposes an upper bound 
$\,m_X^{}\!\!\lesssim\! 27\,(69)$\,GeV
for scalar (fermionic) DM,
which is always satisfied in our models.

\vspace*{-2mm}
\section{\hspace*{-2.5mm}Analyzing Laboratory Constraints}
\label{sec:LabConstraints}
\label{sec:5}
\vspace*{-2mm}

In this section, we proceed to analyze the relevant laboratory constraints
on our inelastic DM models.

\vspace*{-2mm}
\subsection{\hspace*{-2.5mm}Correction to $Z$ Boson Mass} 
\label{sec:SDM}
\label{sec:5.1}
\vspace*{-2mm}

Our model contributes to several electroweak precision observables.
Using Eq.\eqref{eq:boson_masses}, we derive the new correction to $Z$ boson mass:
\vspace*{-2mm}
\beqa
\delta M_Z^{}
=\frac{\,g_X^2v_1^4\,}{~4v_h^2 M_Z^{}\,} \,.
\label{eq:dM_Z}
\eeqa
The $Z$ pole observables have been measured with high precision.
Especially, fitting the SM with the precision data gives the prediction
$\,M_Z^{\text{sm}}\!=\!(91.1884\pm 0.0020)$\,GeV, while
the direct measurement of $Z$ boson mass gives
$\,M_Z^{\text{exp}}\!=(91.1876\pm 0.0021)$\,GeV\,\cite{PDG}.~This
strongly constrains any new physics contribution
to $Z$ mass, $\,\delta M_Z^{} \!< 0.0049$\,GeV at 95\%\,C.L.
Thus, 
we derive an upper bound on the VEV of the Higgs doublet $H_1^{}$,
\begin{equation}
v_1^{}\lesssim 15.2\,\text{GeV} /\!\sqrt{g_X}
\,.
\label{eq:ZmassConstraint}
\end{equation}

\vspace*{-5mm}
\subsection{\hspace*{-2.5mm}Correction to Electron Anomalous\\ 
Magnetic Moment} 
\label{sec:SDM}
\label{sec:5.2}
\vspace*{-2mm}

The $U(1)_X$ interaction also alters the anomalous magnetic moments of electrons.
For $m_{A'}^{}\!\gg m_e^{}$, we derive the one-loop correction to
$\,a_e^{}=\fr{1}{2}(g_e^{}\!-2)$\,:
\vspace*{-1mm}
\beqa
\delta a_e^X \simeq
-\frac{\,g_X^2 q_e^2\,}{\,12\pi^2\,} \frac{m_e^2}{\,m_{A'}^2\,} \,,
\label{eq:e_g-2}
\eeqa
with $\,q_e^{}\!=\!-\fr{1}{2}\,$,
which agrees with \cite{Leveille:1977rc}.
We note that $\,\delta a_e^X <0\,$ is because the mediator $A'$
couples to electrons via right-handed coupling only. 
In contrast, the kinetically mixed dark photon model generally predicts positive correction to $\,a_e^{}$ 
due to its vector-like coupling to the electron. 
For $(g_X^{},\,m_{A'}^{})$ obeying Eq.\eqref{eq:XENON1T_vs}
and for the scalar (fermionic) DM having $U(1)_X$ charge
$\,q_{\text{DM}}^{}\!=\!3\,(3/2)$\,,
we derive the correction to the electron anomalous magnetic moment:
\beqa
\hspace*{-6mm}
\delta a_e^X \,\simeq\, -3.0\,(6.2)\!\times\! 10^{-14}
\!\!\times\!\! \sqrt{m_X/(1\,{\rm GeV})\,}
\,.
\eeqa
These are around the same order of magnitude of 
the current experimental sensitivity of
$\,2\!\times\!10^{-13}$ \cite{PDG}, 
and can be further probed by the future precision measurement of electron magnetic dipole moment especially for the 
fermionic inelastic DM scenario.

\vspace*{-2mm}
\subsection{\hspace*{-2.5mm}Other Constraints on the Higgs Sector} 
\label{sec:SDM}
\label{sec:5.3}
\vspace*{-2mm}

There are additional bounds that can constrain the Higgs sector of
our inelastic DM model.
For instance, the signal strength of the SM-like Higgs boson $h(125$GeV) measured at
the LHC can constrain the mixing between the CP-even neutral component of $H_2^{}$
and the other scalar singlets.
The flavor-dependent feature of our two-Higgs-doublet sector could induce
the flavor-changing processes via Higgs-exchange, so the mass of the
heavy neutral Higgs (mainly from $H_1^{}$ doublet) is constrained by meson mixings.
Since the present study focuses on analyzing the DM and its vector portal
to the SM particles, we note that these constraints can be satisfied
by proper parameter choice in the Higgs sector,
which will be elaborated in Appendix\,\ref{app:B}.

\vspace*{-2mm}
\subsection{\hspace*{-2.5mm}Collider Constraints} 
\label{sec:SDM}
\label{sec:5.4}
\vspace*{-2mm}

Since the mediator $A_\mu'$ couples directly to the right-handed electrons,
our model will receive nontrivial tests by the $e^+e^-$ collider measurements.
(There are discussions on constraining light DM models at lepton colliders
in the literature\,\cite{Fox,Essig:2013vha,Darme:2020ral}.)
The mediator $A_\mu'$ contributes constructively to the cross section of
$\,e^+e^- \!\!\to\!e^+ e^-$.\,
This cross section has been measured by LEP\,\cite{LEP_l_scatt},
with which we infer a bound on our model,
$\,\sqrt{4\pi s\,}/(|q_e| {g}_X^{}) \!>\! 8.6$\,TeV
at $95\%$\,C.L.,$\!$\footnote{%
Hereafter all the quoted experimental bounds are set at 95\%\,C.L.\
unless specified otherwise.}\,
where $\sqrt{s\,}\!\simeq\!200$\,GeV is the LEP collider energy,
and $\,q_e^{}\!=\!-\fr{1}{2}$\, is the $U(1)_X$ charge of $e_R^{}$
in our model.
From this we deduce an upper bound on the $U(1)_X$ gauge coupling,
$\,g_X^{}\!\lesssim 0.16$\,.

\vspace*{1mm}

The DM particles can be pair-produced in $e^+e^-$ collisions through $s$-channel
$A'(Z)$-exchanges and in association with the final state mono-photon,
$e^+e^-\!\to\! XX'\gamma$\,.
The rate of the DM production with mono-photon via $Z$-exchange
is highly suppressed by a coupling factor $v_1^4/v_h^4$
which arises from the gauge boson mixing matrix \eqref{eq:boson_mixing}.
As we have shown in Sec.\,\ref{sec:CosmoConstraints} and Fig.\,\ref{fig:2}-\ref{fig:3},
the $A'$ mass has to be less than a few GeV due to the combined bounds
of realizing the DM relic density and fitting the DM direct detection data.
We find that for the contribution of $A'$  exchange in the case of
$\sqrt{s}\!>\!m_{A'}^{}\!\!>\!2m_X^{}$ at LEP\,\cite{DELPHI},
the cross section of this process has a resonance at
$\,E_\gamma^{}/E_\text{beam}^{}\!\!\sim\!\! 1\!-m_{A'}^2/s$\,.
By fitting the LEP mono-photon data with parameters obeying the relic density bound,
we derive the 95\% exclusion limit on our models,
$\,g_X^{}\!\!\lesssim\! 0.015$\,.
On the other hand,
we find that in the parameter region $\,m_{A'}^{}\!\!<\!2m_X^{}$
and under the DM relic density bound, the contribution to
the LEP mono-photon process is too small to receive constraint,
so this region is always allowed.

The mono-photon searches at low energy $e^+ e^-$ colliders
such as BaBar\,\cite{Aubert:2008as} and Belle-II\,\cite{Abe:2010gxa}
can also place nontrivial bounds on models of light DM\,\cite{Essig:2013vha}.
We summarize these bounds on our model as follows,
according to Ref.\,\cite{Essig:2013vha}.
For the mass region
$\,m_X^{}\!\!\lesssim\! 1$\,GeV and
$\,m_{A'}^{}\!\!\lesssim\! 2m_X^{}$,
the BaBar measurements constrain
$\,g_X^2|q_e^{}| q_{\text{DM}}^{} \!\!<\! O(10^{-2})$\,.
For the on-going Belle-II experiment,
the projected constraint set by null result is
$\,g_X^2q_e^{} q_{\text{DM}^{}}\!\! <\! O(10^{-3})$,
assuming that the backgrounds are ideally known.
As shown in Sec.\,\ref{sec:4}, for this case, combining the constraints from
the DM direct detections and realizing the DM relic density
will require $\,g_X^{}\!\!\lesssim\! 0.016\,$,
which satisfies all the $e^+ e^-$ collider bounds.
For the mass range $\,m_{A'}^{}\!\!>\!2\mX\,$,
the BaBar experiment sets strong constraint on the $A'\!-e^\pm$ coupling.
It is stronger than LEP mono-photon search limit,
due to the much larger production cross section
and the much higher integrated luminosity of BaBar.
Given our charge assignment $\,q_e^{}\!=\!1/2\,$
and the branching fraction of $\,A'\!\!\to\!\! XX'$,
we can derive the corresponding constraint on our model,
$\,g_X^{} \!\!\lesssim 0.003$\,.
Combined with the bounds from the XENON1T electron recoil
and DM relic density,
the BaBar bound on $g_X^{}$ further constrains the DM and mediator masses
$\,m_X\!<\!0.25$\,GeV and $\,m_{A'}^{}\!<\!0.68$\,GeV
for the inelastic scalar DM,
and $\,\mX$ $\!<\!0.21$\,GeV and $\,m_{A'}^{}\!<\!0.55$\,GeV
for the inelastic fermionic DM.
Thus, the $\,m_{A'}^{}\!\!>\!2\mX$\, region is largely excluded.

\vspace*{1mm}

The LHC measurement of $\,Z\!\to\! 4\mu$\, decays\,\cite{LHC} can
constrain the  coupling between $A'_\mu$ and $\mu^\pm$.
This coupling is induced from the neutral gauge boson mixing
matrix \eqref{eq:boson_mixing} and the corresponding Lagrangian term is
\begin{eqnarray}
  \mathcal{L}\supset  \frac{\,g_X^{}v_1^2\,}{\,v_h^2\,}
  \,\bar{\mu}\gamma^\mu\!\left[\(\sin^2\!\theta_W^{}\!-\!\fr{1}{4}\)\!
  +\!\fr{1}{4}\gamma^5\right]\!\!A'_\mu\,\mu \,.
\hspace*{5mm}
\end{eqnarray}
For a vector-type new interaction
$\,g_\text{new}^{}\bar{\mu}\,\gamma^\mu\! A'_\mu \mu$\,
and the small mediator mass $\,m_{A'}^{}\!\!<\!10$\,GeV,
the LHC has placed a bound on its coupling,
$\,g_\text{new}^{}\!\lesssim\! 4.5\!\times\!10^{-3}$
at 95\%\,C.L.
For $\,v_1^2/v_h^2\!=\!10^{-2}$,\,
we convert this LHC bound to a constraint on the $U(1)_X$
coupling of our model,
$\,g_X^{}\!\lesssim\!O(1)$\,,
which is weaker than the combined bound by the DM direct detections and the DM relic density.
The inelastic DM particles of our model can also be directly produced at the LHC
through its coupling to the right-handed quarks $(u_R^{},\,d_R^{})$,
giving raise to mono-photon signals together with the missing $P_T^{}$.
Such signals have been actively searched
by ATLAS\,\cite{ATLAS:2020wzf} and CMS\,\cite{CMS}.
But, these searches lose sensitivity for
$\,g_X^{}\!\!<\! O(0.1)$ \cite{Buchmueller:2014yoa}
and does not constrain the cosmologically favored parameter space
as discussed in Sec.\,\ref{sec:4}.

\vspace*{-2mm}
\section{\hspace*{-2.5mm}Conclusions}
\label{sec:6}
\label{sec:Conclusion}
\vspace*{-2mm}

In this work, we proposed a new candidate of GeV scale inelastic 
dark matter (DM), which can be either scalars or fermions.
For this we constructed anomaly-free and 
renormalizable inelastic DM models
under a new $U(1)_X$ gauge symmetry with dark photon mediator
(without assuming kinetic mixing)
and with scalar or fermionic DM particles
(cf.\ Table\,\ref{tab:1}).
We realized the natural $O(\text{keV})$ mass-splitting for the
inelastic DM by a scalar seesaw mechanism. 
Our model resolved the recent XENON1T anomaly 
in the electron recoil detection\,\cite{Aprile:2020tmw}.
It is highly predictive and testable. We further analyzed the nontrivial bounds
from the nuclear recoil detection (including Migdal effect)
by the XENON1T\,\cite{Aprile:2019jmx}, CRESST-III\,\cite{Abdelhameed:2019hmk},
CDEX-1B\,\cite{CDEX-1B}, and DarkSide-50\,\cite{DarkS}
experiments with low recoil energy thresholds.
Combining the constraints from both the electron recoil 
and nuclear recoil detections, 
we identified the viable parameter space and
predicted the inelastic DM mass $\lesssim\!\!1.5\,$GeV,\,
as shown in Fig.\,\ref{fig:1}(b).
Then, we derived the viable parameter space in
Fig.\,\ref{fig:2} for the scalar inelastic DM and
Fig.\,\ref{fig:3} for the fermionic inelastic DM,
under the constraints by the DM relic abundance,
the lifetime of the heavier DM component,
the electroweak precision tests, and the collider searches.
The upcoming DM direct detection experiments by
the PandaX-4T\,\cite{PandaX4T}, LZ\,\cite{LZ}, and XENONnT\,\cite{XENONnT}
collaborations will provide decisive probes of
our GeV scale inelastic DM candidate.

%
\vspace*{4mm}
\noindent
{\bf\large Acknowledgements}
\\[0.8mm]
We thank Jianglai Liu, Qian Yue, and Ning Zhou for useful discussions
on DM direct detections.
This research was supported in part by National NSF of China
(under grants No.\,11835005 and No.\,12175136), by National Key R\,\&\,D Program
of China (under grant No.\,2017YFA0402204), and by the CAS Center for Excellence
in Particle Physics (CCEPP).  It was also supported in part
by Office of Science and Technology, Shanghai Municipal Government.


\appendix

\vspace*{4mm}
\begin{center}
{\bf\large Appendix}		
\end{center}

\vspace*{-7mm}

\section{\hspace*{-2.5mm}Improved Constraint of
Migdal Effect\\ on the Inelastic Dark Matter}
\label{sec:APD:inelastic_migdal_est}
\label{app:A}

\vspace*{-2mm}

In this Appendix, we make an estimate of the improved constraint on the inelastic DM by using a 
scaling relation of the Migdal effect process. 

\vspace*{1mm}

We begin by recalling the formulation of Migdal process for the direct detection of a single component dark matter \cite{Migdal}. The differential rate of electron recoil (ER) in the Migdal process 
per unit target mass is given by
\begin{eqnarray}
\hspace*{-2mm}
\frac{{\rm d}R}{{\rm d}E_{\rm ER}\,} \simeq  
\int\!\! {\rm d}E_{\rm NR} {\rm d}v\!
\Bigg[\!
\frac{{\rm d}^2 R_{\rm N}}{\,{\rm d}E_{\rm NR} {\rm d}v\,}\!
\sum_{n,\ell} \!\frac{1}{2\pi}
\frac{{\rm d}P^c_{q_e}\!(n,\ell\!\to\!E_{e})}
{\,{\rm{d}}E_{\rm ER}}  
\!\Bigg]\!,~
\nn\\
\label{eq:migdal_master}
\end{eqnarray} 
where the differential rate of the nuclear recoil 
per unit target mass is
\begin{equation}
\frac{\d R_{\rm N}}{\,{\d}E_{\rm NR}\d v\,}
\,\simeq\,
\frac{\rho_{X}^{}\sigma_{\!A}^{}}{\,2\,m_{\!X}^{}\mu_{\!A}^2\,}
\frac{f(v)}{v}\,.\label{eq:NR_rate}
\end{equation}
In the above, $\rho_X^{}$ is the local DM density, 
$\mu_{\!A}^{}$ is the DM-atom reduced mass, and  $\,\sigma_{\!A}^{}\!\equiv\!\frac{\left|F_{A}\right|^{2}
\left|{\cal M}\right|^{2}}{\,(m_{A}+m_{X})^{2}\,}$
is a parameterization of atomic form factor $F_{\!A}^{}$ 
and the amplitude. It equals the DM-nuclear cross section
for elastic scattering, but not for the inelastic case. 
The function $f(v)$ is the distribution 
of the local DM velocity $v$\,. 
The $P^c_{\!q_{e}}(n,\ell\!\to\! E_{e})$ is the probability of exciting an $(n,\ell)$ electron with ionization energy 
$E_{n,\ell}$ to an unbounded electron with energy deposit $\,E_{e}\!=\!E_{\rm ER}^{}\!-\!E_{n,\ell}$\,.\ 
The momentum $\,q_e^{}\!=\!m_e v_{\!A}^{}$,\, 
where $v_{\!A}^{}$ 
is the atom velocity after scattering in the lab frame. 
We can further write
\begin{equation}
\hspace*{-2mm}
\frac{1}{2\pi}
\frac{\d P_{q_e}^{}\!(n,\ell\!\to\! E_{e})\,}
{{\d}E_{e}}
\equiv\, q_{e}^{2}A_{n,\ell}^{}(E_{e})\,, 
\label{eq:Prob_Dist}
\end{equation}
%
where 
$\,q_e^2\!=\!{\,2m_{e}^{2}E_{\rm NR}}/{m_{\!A}^{}}$\,.\
Thus, $A_{n,\ell}^{}(E_{e})$\, is only a function of $E_e$\,\cite{Migdal}, 
but independent of $E_{\rm NR}$.

\vspace*{1mm}

In general, the detected energy $E_{\rm det}^{}$ receives contribution from both nuclear and electron recoil, 
$E_{{\rm det}}=E_{\rm ER}+{\cal L}E_{\rm NR}$,\, 
with $\,{\cal L}\approx 0.15$\, 
as the quenching factor of nuclear recoil. 
For small 
$\,\mX\!\lesssim\!1\,$GeV and 
$\delta=O({\rm keV})$,\, 
the nuclear recoil energy 
$\,E_{\rm NR}^{}\!\lesssim\!\frac{\mu_{\!A}^{}}{\,m_{\!A}^{}\,}
(\frac{1}{2}\mu_{\!A}^{}v^{2}+\delta)$\,
is rather small, so its contribution to the detected energy 
can be neglected in the following estimate for
the sensitive range of XENON1T with  
$\,E_{{\rm det}}\!\simeq\! E_{\rm ER}\!\gtrsim0.1\,$keV.

\vspace*{1mm}

For the spin-independent interaction, the
DM-nuclear cross sections $\,\sigma_{\!A}^{}$\, 
is independent of the nuclear recoil energy. 
We can perform the integration
over $E_{\rm NR}^{}$ analytically
in Eq.\eqref{eq:migdal_master}, and obtain the result:%
\footnote{We have evaluated Eq.\eqref{eq:IntER} numerically and confirmed its good agreement with most curves in Figs.2-3 of Ref.\,\cite{Bell:2021zkr} for 
$E_{\rm ER}^{}\!\!\gtrsim\! 0.1$\,keV. 
We obtained a less suppressed exothermic scattering spectrum 
in the range $E_{\rm ER}^{}\!\gtrsim\! 1$\,keV 
for the DM of $7$\,MeV mass 
in comparison with the green curve in Fig.3 
of Ref.\,\cite{Bell:2021zkr}. 
We have confirmed this with Jayden Newstead 
(the coauthor of Ref.\,\cite{Bell:2021zkr})
and this discrepancy can be traced back to a minor coding error 
in producing the curves of Ref.\,\cite{Bell:2021zkr}. 
We thank Jayden Newstead for clarification and for sharing his 
code of Ref.\,\cite{Bell:2021zkr}.}
\begin{align}
\hspace*{-2mm}
\frac{\d R}{\d E_{\rm ER}} \!=&\,\frac{1}{\,m_{\!A}^{}}
\frac{\,\rho_{X}^{}\sigma_{\!A}^{}m_e^2\,\mu_{A}^{2}\,}{\mX m_A^2}
A_{n,\ell}^{}(E_{\rm ER}^{}\!-E_{n,\ell}^{})
\label{eq:IntER}\\
\hspace*{-2mm}
& 
\times\! 2\!\int_{v_{\min}}^{}\hspace*{-5.5mm}
{\d}v f(v)v
\sqrt{\!1\!+\!\frac{\,2(\delta\!-\!E_{\rm ER})}{\mu_{\!A}^{}v^{2}}}
\!\(\!\!\frac{\,\delta\!-\!E_{\rm ER}\!+\!\mu_{\!A}^{}v^2\,}
{\mu_{\!A}^{}}\!\!\)\!.
\nn
\end{align}
For a fixed $E_{\rm ER}$, the minimum velocity 
$v_{\min}^{}(E_{\rm ER})$ to excite the electron is given by
\begin{equation}
v_{{\rm min}}(E_{\rm ER})^2=
{\max}\!\left[\frac{\,2(E_{\rm ER}\!-\delta)\,}{\mu_{A}},
\,0\,\right]\!.
\end{equation}
$v_{\min}^{}(E_{\rm ER})\!=\!0\,$ for 
$\,E_{\rm ER}\!<\!\delta\,$. 
For $\,\delta\!\simeq\!2.8\,$keV considered in the present study, 
the major part of the recoil spectrum of the Migdal process 
lies within 
$\,E_{\rm ER}\!\lesssim\!1\,$keV\,\cite{Bell:2021zkr}. 
As an estimate, we set $v$ as the most probable velocity $v_{0}\approx0.77\times10^{-3}$ of the local DM.
Thus, we obtain an approximate differential spectrum for $\mu_A\!\approx\! m_X$:
\begin{equation}
\hspace*{-3mm}
\frac{\d R}{\d E_{\rm ER}}\propto\,
\frac{\mu_{A}^{2}}{\mu_{N}^{2}}\sigma_{\!N}^{}
\sqrt{1\!+\!\frac{\,2(\delta\!-\!E_{\rm ER})\,}
{\mu_{\!A}^{}v_0^{2}}}
\(\!
{\,\delta\!-\!E_{\rm ER}\!+\mu_{\!A}^{}v_0^{2}\,}
\)\!.
\label{eq:Migdal_scaling}
\end{equation}
Here, $\mu_{\!N}^{}$ is the DM-nucleon reduced mass 
and $\sigma_{\!N}^{}$ the DM-nucleon cross section. 
To utilize the direct detection results, we have adopted the convention that DM coupled equally to neutron and proton in deriving Eq.\eqref{eq:Migdal_scaling}, and thus\,%
\footnote{The inelastic DM candidate studied in the main text 
couples to neutron and proton with opposite charges. 
This requires additional treatment as we have shown in Sec.\,\ref{sec:3} of our main text.}
$\,\sigma_{\!A}^{}/\sigma_{\!N}^{}
\!=\!A^{2}\mu_{\!A}^2/\mu_{\!N}^{2}$\,.  
For the light DM, the difference 
$\,\delta-E_{\rm ER}\!\gg\!\mu_{\!A}^{}v^2$\, and
the recoil spectrum scales as $\sqrt{1/\mX\,}$\,. 

\vspace*{1mm} 

So far in the literature the constraints by including Migdal effect  
are only given for elastic DM and for inelastic DM 
with a few special inputs of $\delta$ \cite{Bell:2021zkr}. 
We estimate the constraint on the (exothermic) inelastic DM 
candidate with $\delta\!=2.8\,$keV (cf.\ main text) 
based on the following observations.\  
Ref.\cite{Bell:2021zkr} shows
that the detectable side of the spectrums are similar 
for elastic and inelastic DM with parameters 
$(\delta/{\rm keV},\,\mX/{\rm GeV},\,
\sigma_{\!N}^{}/(10^{-40}{\rm cm^2}))\!=\!
(0,\,2,\,1)$ and $(4,\,0.5,\,0.65)$, respectively.\ 
According to the scaling relation of
Eq.\eqref{eq:Migdal_scaling}, 
for $\,\delta\!=\!2.8\,$keV, 
the following input parameters 
$(\mX/{\rm GeV},\,
\sigma_{\!N}^{}/(10^{-40}{\rm cm^{2}}))\!=\!(2,\,0.41)$,
$(1,\,0.83)$, $(0.5,\,1.20)$, and $(0.1,\,1.13)$\, 
give rise to the electron recoil spectra
which are similar to that of 
$(\delta/{\rm keV},\,\mX/{\rm GeV},\,\sigma_{\!N}^{}/(10^{-40}
{\rm cm^{2}}))\!=\!(4,\,0.5,\,0.65)$,
and thus are also similar to that of elastic DM with $(m_{X}/{\rm GeV},\sigma_{N}/(10^{-40}{\rm cm^{2}}))=(2,1)$. Since the 
constraint of XENON1T Migdal effect is already given 
for the elastic DM cross section\,\cite{Aprile:2019jmx}, 
we can estimate the constraint on the inelastic DM by 
rescaling the elastic DM constraint according to the parameters 
that gives rise to similar spectrum. 
We show our estimate by the purple dashed curve 
in Fig.\ref{fig:1}(b).

\vspace*{-1mm}
\section{\hspace*{-2.5mm}Constraints on the Higgs Sector}
\label{app:B}
\label{sec:APD:HiggsConstraints}

In this Appendix, we present the constraints on the Higgs sector of our model.
For convenience, we denote the CP-even neutral components of
$(H_1^{},\,H_2^{},\,S')$ as
$(h_1^{},\,h_2^{},\,h_{S'}^{})$, respectively.
We consider the case of $\,M_{h_1}^2\!\gg\! M_{h_2}^2\,$ and
$\,v_1^2\!\ll\! v_2^2$\,,\,
so the observed Higgs boson $h(125\text{GeV})$ mainly contains the
$h_2^{}$ state. To realize these conditions,
we consider the relevant part of the scalar potential,
\begin{eqnarray}
V &\!\!\supset\!\!&
M_{H_1}^2|H_1|^2\!
-M_{H_2}^2|H_2|^2\!
-M_{S'}^2|S'|^2\!
- M' H_1^\dagger H_2 S'
\hspace*{5mm}
\nonumber\\
&&
\!+ \lambda_1 |H_1|^4
\!+ \lambda_2 |H_2|^4
\!+ \lambda_3 |H_1|^2|H_2|^2
\!+ \lambda_{S'} |S'|^4
\hspace*{5mm}
\nonumber\\
&&
+ \lambda_{H_1 S'}^{} |H_1|^2|S'|^2
+ \lambda_{H_2 S'}^{} |H_2|^2|S'|^2,
\label{eq:VV}
\end{eqnarray}
where we only list terms relevant to $H_1$, $H_2$ and $S'$.\,
The cubic term $\,H_1^\dagger H_2 S'$ will ensure nonzero mass
of the pseudoscalars.
In Eq.\eqref{eq:VV}, we take all the mass parameters and the quartic couplings
be positive. We choose the Higgs doublet $H_1^{}$ to have a 
positive mass term, so its mass $M_{H_1}^{}$ can be naturally large
and $M_{H_1}^2\!\!\gg v_2^2,v_{S'}^2$.
Thus, we deduce the VEV of $H_1^{}$,
\vspace*{-2mm}
\beqa
v_1^{} \simeq \frac{\,v_2^{}v_{S'}^{}M'\,}{\,M_{H_1}^2\,}\,.
\\[-4mm]
\nn
\label{eq:v1_2HDM}
\eeqa
We see that requiring
$\,M_{H_1}^2 \!\!\gg\! M' v_{S'}^{}$ can realize
$\,v_1^2\ll v_2^2$\,.
The LHC ATLAS and CMS experiments
have measured the signal strength of the Higgs boson $\,h$(125GeV)
for various channels, defined as
$\,\mu_h^{} \equiv \left< \sigma_h^{}\!\cdot \text{BR}\right>_\text{obs} /\!
 \left< \sigma_h^{}\!\cdot \text{BR}\right>_\text{SM}$,
with $\,\sigma_h^{}$ the $h$ production cross section
and BR the decay branching fraction of a given channel.
The most precisely measured decay channels are $\gamma\gamma$ and $W W^*$
which are consistent with the SM prediction ($\mu_h^{}\!\!=\!1$)
to $10\%$ level\,\cite{PDG}.
This constrains the mixings of the SM-like Higgs boson
$h_2^{}$ with $h_{S'}^{}$ and $h_1^{}$ down to $10\%$ level.
Note that for $\,v_1^2\!\ll\! v_2^2$\,,
the $h_2^{}$-$h_1^{}$ mixing is mainly generated by the cubic term
$\,- M' H_1^\dagger H_2 S'$,\,
and the contributions from the mixed quartic terms are suppressed
by $v_1^{}/M'$.
Thus, for $\,M_{H_1}^2\!\gg\! M_{H_2}^2$, the $h_2^{}$-$h_1^{}$
mixing is mainly determined by
$|M'v_{S'}^{}/M_{H_1}^2|\!\simeq\! v_1^{}/v_2^{}$\,.
With this condition, we may choose the VEV ratio
\beqa
v_1^{} / v_2^{} \lesssim 0.1 \,.
\label{eq:v_ratio}
\eeqa
The small $h_2^{}$-$h_{S'}^{}$ mixing can be induced
by the contributions from the cubic and quartic terms
with opposite signs. As an estimate, we ignore the small mixing of
the heavy $h_1$ with the lighter states $h_2^{}$ or $h_{S'}^{}$
and obtain,
\beqa
\left|\sin\theta_{h_2 h_{S'}}\right|
\simeq
\left|
\frac
{\,\lambda_{H_2\! S'}^{} v_2^{} v_{S'}^{}\!-\!M' v_1^{}\,}
{M_{h_{S'}}^2 \!\!-\! M_h^2}
\right|\ll 1
\,,
\label{eq:h2-hsp-mixing}
\eeqa
where $M_h^{}\!\!\simeq\! 125$GeV and $M_{h_{S'}}^{}$ are the
mass-eigenvalues of the CP-even neutral Higgs states $h$ and $h_{S'}^{}$.

Another constraint on the heavy Higgs mass $M_{H_1}$
comes from the flavor-changing processes mediated by the heavy scalars.
The general Yukawa interactions for the quark sector take the following form,
\begin{subequations}
\vspace*{-1.5mm}
\begin{eqnarray}
\hspace*{-6mm}
{\cal L}
&\!=\!&	-\sum_{i=1}^{3}\(y_{u}^{i1}\bar{Q}_{Li}\tilde{H}_{1}u_{R1}\!
+y_{d}^{i1}\bar{Q}_{Li}H_{1}d_{R1}\)
\nn\\
\hspace*{-6mm}
&& \ \ \ -\sum_{i=1,2,3}^{j=2,3}\!\!
	\(y_{u}^{ij}\bar{Q}_{Li}\tilde{H}_{2}u_{Rj}\!+y_{d}^{ij}\bar{Q}_{Li}H_{2}d_{Rj}\)
\\
\hspace*{-6mm}
&\!\!\equiv\!&
-\bar{Q}_{L}^{i}\!\!\!\left\lgroup\!
\vec{\bf y}_{\bf\!u}\tilde{H}_{\!1}, \mathbf{\hat{y}_{\!u}}
\tilde{H}_{2}\!\right\rgroup_{\!\!ij}\!\!\!u_{Rj}^{}\! -\! \bar{Q}_{L}^{i}\!
\!\!\left\lgroup{\vec{\bf y}_{\bf\!d}}H_{\!1}^{},
\mathbf{\hat{y}_{\!d}}H_2^{}\!\right\rgroup_{\!\!ij}\!\!\!d_{Rj}^{}
\hspace*{5mm}
\end{eqnarray}
\end{subequations}
where we denote $\tilde{H}_{i}\!=\!{\rm i}\sigma_{2}H_{i}^{*}$.
For convenience, in the last line, we have decomposed the $3\!\times\!3$
Yukawa matrix
$\mathbf{y^{u,d}}=(\vec{\bf y}^{}_{\bf\!u,d},\,\mathbf{\hat{y}_{\!u,d}^{}})_{ij}^{}$.
Here, $\vec{\bf y}_{\bf\!u,d}^{}$ are $3\!\times\!1$ matrices of Yukawa
couplings to $H_{1}^{}$ and $\mathbf{\hat{y}_{u,d}^{}}$ are $3\!\times\!2$ matrices
of Yukawa couplings to $H_2^{}$\,. The quarks acquire masses
via Yukawa interactions with $H_{1}$ and $H_{2}$ taking their VEVs.
The mass eigenstates are obtained by chiral rotations,
\begin{subequations}
\addtolength{\jot}{0.3em}
\begin{align}
&	\mathbf{u_{L}'} = U^{u}_L \mathbf{u_{L}}\,,\hspace*{7mm}
	\mathbf{d_{L}'} = U^{d}_L \mathbf{d_{L}}\,,
\\
&	\mathbf{u_{R}'} = U^{u\dagger}_R \mathbf{u_{R}}\,,\hspace*{5mm}
	\mathbf{d_{R}'} = U^{d\dagger}_R \mathbf{d_{R}}\,,
\hspace*{7mm}
\end{align}
\end{subequations}
where $\bf u$ and $\bf d$ are vectors in flavor space denoting the 3 family of quarks,
and $U^{u,d}_{L,R}$ are unitary rotation matrices.
The quark mass matrices are diagonalized as
\beqs
\beqa
U^{u\dagger}_L\!\!
\left\lgroup\vec{\bf y}_{\bf\!u}^{}v_1^{},\,
\hat{\bf y}_{\bf\!u}^{}v_2^{}\!\right\rgroup
\!\!U^{u}_R
&\!=\!& v_2^{}\,\mathbf{y}^{\text{diag}}_{\bf\!u}\,,
\\
U^{d\dagger}_L\!\!
\left\lgroup\vec{\bf y}_{\bf\!d}^{}v_1^{},\,
\hat{\bf y}_{\bf\!d}^{}v_2^{}\!\right\rgroup
\!\!U^{d}_R
&\!=\!& v_2^{}\,\mathbf{y}^{\text{diag}}_{\bf\!d}\,,
\eeqa
\eeqs
where $\,v_2^{}\mathbf{y}^{\text{diag}}_{\bf\!u(d)}$\,
is the $3\!\times\!3$ diagonal mass matrix for up-type (down-type) quarks,
whose diagonal elements give the measured quark masses.
In the mass-eigenbasis, the Yukawa interactions
for the up-type quarks become:
\vspace*{-1mm}
\begin{subequations}
\begin{eqnarray}
{\cal L}  &=&
-\mathbf{\bar{u}'_L}U^{u\dagger}_L\!\!
\left\lgroup\!\vec{\bf y}_{\bf\!u}^{}\tilde{H}_{1}^0,\,
\mathbf{\hat{y}_{\!u}^{}}\tilde{H}_{2}^0\!\right\rgroup\!\! U^{u}_R \mathbf{u'_R}
\\
&=&
-\mathbf{\bar{u}'_{L}}U^{u\dagger}_L
		\!\!\left\lgroup\!
		\vec{\bf y}_{\bf\!u}\!\!\(\!\tilde{H}_{1}^0\!
		-\!\frac{\,v_1\,}{\,v_2\,}\tilde{H}_{2}^0\!\)\!,\, ({\bf 0})
		\!\right\rgroup\!\!
		U^{u}_R \mathbf{u'_{R}}  \hspace*{10mm}
\nonumber\\
&&
-\mathbf{\bar{u}'_{L}}
		\mathbf{y}^{\text{diag}}_{\bf\!u} \tilde{H}_{2}^0
		\mathbf{u'_{R}} \,,
\label{eq:FC}
\end{eqnarray}
\end{subequations}
where $\tilde{H}_1^0$ and $\tilde{H}_2^0$ denote the neutral components
of the Higgs doublets $\tilde{H}_1^{}$ and $\tilde{H}_2^{}$, respectively.
Similarly, we can deduce the Yukawa interactions for the mass-eigentstates
of down-type quarks under the replacement
$\,\mathbf{u}\!\to\!\mathbf{d}$\, and
$\,\tilde{H}_i\!\to\! {H}_i$\,.
In Eq.\eqref{eq:FC},
$({\bf 0})$ denotes a $3\!\times\! 2$ matrix in which all elements vanish.
The first term in Eq.\eqref{eq:FC} would induce flavor-changing processes
if the flavor mixing matrices $U_{L,R}^u$ take arbitrary pattern,
and in this case it will receive strong constraints by the meson mixings.
It is known that for well-motivated scenarios of flavor mixing
such constraints can be much reduced.
For instance, we may consider Cheng-Sher-like
ansatz\,\cite{2HDM}\cite{Cheng:1987rs}
on the flavor-changing Yukawa couplings,
\beqa
\xi_{u,d}^{ij}	= \bar{\xi}_{u,d}^{ij}\!\times\!
\(\!\!\sqrt{m_{u,d}^i m_{u,d}^j\,} \!\Big/\!v_h^{}\!\!\)\!,
\eeqa
where $(i,\,j)$ are family indices and the coupling coefficients
$\,\bar{\xi}_{u,d}^{ij}\,$
can be naturally around
$\,\bar{\xi}_{u,d}^{ij}\!=\!O(0.1\!-\!1)\,$.
Thus, we set
\beqa
	U^{u,d\dagger}_L
	\!\!\left\lgroup\!
	\vec{\bf y}_{\bf\!u,d},\,({\bf 0})
	\!\right\rgroup\!\!
	U^{u,d}_R =\, \xi_{u,d}^{}\,.
\eeqa
Then, the flavor-changing process between the $i$-th and $j$-th families
is characterized by the new physics (NP) scale,
\beqa
\hspace*{-6mm}
\Lambda_{\rm NP} \,=\,
	\min\!\(\!M_{h_1}^{}\!,\, \frac{v_2}{v_1}M_{h_2}^{}\!\)
	\!\frac{\bar{\xi}_{u,d}^{ij}v_h^{}}{\sqrt{m_{u,d}^i m_{u,d}^j\,}\,}\,,
\eeqa
where the $M_{h_1}^{}\!\!\!=O(5\!-\!10)$TeV 
is the mass of the heavy neutral Higgs state $h_1^{}$
and $M_{h_2}^{}\!\!\!\simeq\!\! 125$\,GeV
is the SM-like light Higgs boson $h_2^{}$\,.
For instance, the measurements of $K$-$\bar{K}$ mixing constrain
$\Lambda_{\rm NP}^{}\!\gtrsim\! 5\!\times\!10^5$\,TeV\,\cite{Alpigiani:2017lpj}.
With this bound and for a natural coupling coefficient
$\,\bar{\xi}_{d}^{12}\!=\!O(0.1)$,\,
we obtain the limits on our Higgs sector,
$\,M_{h_1}^{} \!\!\gtrsim\! 6$\,TeV
and $v_1^{}\!\lesssim\! 3.6$\,GeV.
The $U(1)_X$ gauge boson $A'_\mu$ also mediates flavor-violating process
since it only couples to the right-handed quarks and leptons in the first family.
The ratio $\,m_{A'}^{}/{g_X}^{}$ is constrained by fitting the XENON1T data
as in Eq.\eqref{eq:XENON1T_vs}.
So the constraints from the meson mixing measurements could be avoided
by requiring the right-handed mass-eigenstates
$u'_R$ and $d'_R$ to be mainly aligned with the gauge-eigenstates
$u_R^{}$ and $d_R^{}$.
In addition, we note that the flavor-violating effects can also occur
in the lepton sector and induces flavor-violating leptonic decay channels
for the SM-like light Higgs boson $h_2^{}$\,.
For instance, this leads to the interesting decay channel
$\,h_2^{}\!\to\! \mu^\pm e^\mp$,\,
which can be searched by analyzing the current LHC Run-2 data\,\cite{Aad:2019ojw}.
The upcoming LHC Run-3 and HL-LHC runs will have strong potential
to discover this channel.

\vspace*{0.6mm}

\vspace{-3mm}
%


\begin{thebibliography}{99}
	
	
	\bibitem{Aprile:2020yad}
	E.\ Aprile {\it et al.} [XENON Collaboration],
	``Energy resolution and linearity in the keV to MeV range measured in XENON1T",
	Eur.\ Phys.\ J.\ C\,80 (2020) 785, no.8
	[arXiv:2003.03825 [physics.ins-det]].
	
	
	\bibitem{Aprile:2020tmw}
	E.\ Aprile {\it et al.} [XENON Collaboration],
	``Observation of Excess Electronic Recoil Events in XENON1T",
	Phys.\ Rev.\ D\,102 (2020) 072004, no.7,
	[arXiv:2006.09721 [hep-ex]].
	
	
	\bibitem{PandaX2}
	X.\ Zhou {\it et al.} [PandaX-II Collaboration],
	``A search for solar axions and anomalous neutrino magnetic moment
	with the complete PandaX-II data'',
	Chin.\ Phys.\ Lett.\ (Express) 38 (2021) 011301, no.1 
	[arXiv:2008.06485 [hep-ex]].
	
	
	\bibitem{Tritium}
	A.~E.~Robinson,
	arXiv:2006.13278 [hep-ex].
	
	
	\bibitem{Strumia}
	K.~Kannike, M.~Raidal, H.~Veermae, A.~Strumia, and D.~Teresi,
	``Dark Matter and the XENON1T electron recoil excess'',
	Phys.\ Rev.\ D\,102 (2020) 095002, no.9
	[arXiv:2006.10735 [hep-ph]].
	
	
	\bibitem{others}
	%
	L.~Di Luzio, M.~Fedele, M.~Giannotti, F.~Mescia and E.~Nardi,
	arXiv:2006.12487 [hep-ph].
	%
	F.\ Takahashi, M.\ Yamada and W.\ Yin,
	arXiv:2006.10035 [hep-ph];
	%
	G.\ Alonso-Alvarez, F.\ Ertas, J.\ Jaeckel, F.\ Kahlhoefer, and L.\ J.\ and Thormaehlen,
	arXiv:2006.11243 [hep-ph];
	%
	C.\ Boehm, D.\ G.\ Cerdeno, M.\ Fairbairn, P.\,A.\,N.\ Machado, and A.\ C.\ Vincent,
	arXiv:2006.11250 [hep-ph];
	%
	D.\ A.\ Sierra, V.\ De\,Romeri, L.\,J.\ Flores, and D.\,K.\ Papoulias,
	arXiv:2006.12457 [hep-ph];
	%
	B.\ Fornal, P.\ Sandick, J.\ Shu, M.\ Su,  and Y.\ Zhao,
	arXiv:2006.11264 [hep-ph];
	L.~Su, W.~Wang, L.~Wu, J.~M.~Yang and B.~Zhu,
	arXiv:2006.11837 [hep-ph];
	Y.\ Chen, J.\ Shu,  X.\ Xue,  G.\ Yuan, and Q.\ Yuan,
	arXiv:2006.12447 [hep-ph];
	Q.\ H.\ Cao, R.\ Ding, and Q.\ F.\ Xiang,
	arXiv:2006.12767 [hep-ph];
	%
	H.\ Alhazmi, D.\ Kim, K.\ Kong, G.\ Mohlabeng, J.\ C.\ Park and S.\ Shin,
	arXiv:2006.16252 [hep-ph];
	%
	Y.\ Jho, J.\ C.\ Park, S.\ C.\ Park, and P.\ Y.\ Tseng,
	arXiv:2006.13910 [hep-ph];
	S.\ Chigusa, M.\ Endo and K.\ Kohri, arXiv:2007.01663 [hep-ph].
	%
	J.\ Smirnov and J.\ F.\ Beacom,
	arXiv:2002.04038 [hep-ph];
	%
	A.\ Bally,  S.\ Jana, and A.\ Trautner,
	arXiv:2006.11919 [hep-ph];
	%
	M.\ Du, J.\ Liang, Z.\ Liu, V.\ Q.\ Tran,  and Y.\ Xue,
	arXiv:2006.11949 [hep-ph];
	%
	D.\ Aristizabal Sierra, V.\ De Romeri, L.\ J.\ Flores,\,and D.\ K.\ Papoulias,
	arXiv:2006.12457 [hep-ph];
	%
	N.\ F.\ Bell,  J.\ B.\ Dent, B.\ Dutta, S.\ Ghosh, J.\ Kumar, and J.\ L.\ Newstead,
	arXiv:2006.12461 [hep-ph];
	%
	G.\ Paz, A.\ A.\ Petrov, M.\ Tammaro, and J.\ Zupan,
	arXiv:2006.12462 [hep-ph];
	%
	J.\ Buch, M.\ A.\ Buen-Abad, J.\ Fan, and J.\ S.\ C.\ Leung,
	arXiv:2006.12488 [hep-ph];
	%
	K.\ U.\ Dey, T.\ N.\ Maity, and T.\ S.\ Ray,
	arXiv:2006.12529 [hep-ph];
	%
	A.~N.~Khan,
	arXiv:2006.12887 [hep-ph];
	%
	K.~Nakayama and Y.~Tang,
	arXiv:2006.13159 [hep-ph];
	%
	L.~Zu, G.~W.~Yuan, L.~Feng and Y.~Z.~Fan,
	arXiv:2006.14577 [hep-ph];
	%
	M.\ Lindner, Y.\ Mambrini, T.\ B.\ Melo, and F.\ S.\ Queiroz,
	arXiv:2006.14590 [hep-ph].
	%
	W.\ DeRocco, P.\ W.\ Graham, S.\ Rajendran,
	arXiv:2006.15112 [hep-ph];
	%
	M.\ Chala and A.\ Titov,
	arXiv:2006.14596 [hep-ph];
	%
	C.\ Gao, J.\ Liu, L.\,T.\ Wang, X.\,P.\ Wang, W.\ Xue, and Y.\,M.\ Zhong,
	arXiv:2006.14598 [hep-ph];
	%
	J.\ B.\ Dent, B.\ Dutta, J.\ L.\ Newstead, and A.\ Thompson,
	arXiv:2006.15118 [hep-ph];
	%
	P.\ Ko and Y.\ Tang, arXiv:2006.15822 [hep-ph];
	%
	W.\ Chao, Y.\ Gao, and M.\ Jin, arXiv:2006.16145 [hep-ph];
	%
	L. Delle\,Rose, G.\ Hütsi, C.\ Marzo, and L.\ Marzola,
	arXiv:2006.16078 [hep-ph];
	%
	B.\ Bhattacherjee and R.\ Sengupta,
	arXiv:2006.16172 [hep-ph];
	%
	Y.\ Gao and T.\ Li,
	arXiv:2006.16192 [hep-ph];
    %
    J.\ Sun and X.\,G.\ He,
    arXiv:2006.16931 [hep-ph];
	%
	M.\ Szydagis, C.\ Levy, G.\ M.\ Blockinger, A.\ Kamaha,
    N.\ Parveen, and G.\ R.\ C.\ Rischbieter,
	arXiv: 2007.00528 [hep-ex];
	%
	T.\ Li, arXiv:2007.00874 [hep-ph];
	%
	O.\ G.\ Miranda, D.\ K.\ Papoulias, M.\ Tortola, and J.\,W.\,F.\ Valle,
	arXiv:2007.01765 [hep-ph];
	%
	K.\ Benakli, C.\ Branchina and
	G.\ Lafforgue-Marmet,
	arXiv:2007.02655 [hep-ph];
	%
	N.\ Okada, S.\ Okada, D.\ Raut, and Q.\ Shafi, arXiv:2007.02898 [hep-ph];
	%
	J.\ Davighi, M.\ McCullough, and J.\ Tooby-Smith,
	arXiv:2007.03662 [hep-ph];
	%
	P.\ Athron,\ \textit{et al.},
	arXiv:2007.05517 [astro-ph.CO];
	%
	G.~Arcadi, A.~Bally, F.~Goertz, K.~Tame-Narvaez, V.~Tenorth and S.~Vogl,
	arXiv:2007.08500 [hep-ph];
	%
	C.\ Han, M.\ L.\ Lopez-Ibanez, A.\ Melis, O.\ Vives and J.\ M.\ Yang,
	arXiv:2007.08834 [hep-ph];
	%
	Y.\ Ema, F.\ Sala and R.\ Sato,
	arXiv:2007.09105 [hep-ph];
	%
	J.\ Kim, T.\ Nomura and H.\ Okada,
	arXiv:2007.09894 [hep-ph];
	%
	J.\ Cao, X.\ Du, Z.\ Li, F.\ Wang and Y.\ Zhang,
	arXiv:2007.09981 [hep-ph];
	%
	D.\ Borah, S.\ Mahapatra, D.\ Nanda and N.\ Sahu,
	arXiv:2007.10754 [hep-ph].
	%
	S.~Karmakar and S.~Pandey,
	arXiv:2007.11892 [hep-ph];
	%
	S.~Khan,
	arXiv:2007.13008 [hep-ph];
	%
	S.\ Shakeri, F.\ Hajkarim, and S.\ S.\ Xue,
	arXiv:2008.05029 [hep-ph];
	%
	R.\ G.\ Cai, S.\ Sun, B.\ Zhang, and Y.\ L.\ Zhang,
	arXiv:2009.02315 [hep-ph];
	%
	R.\ Foot, arXiv:2011.02590 [hep-ph].
	%
	Y.~Farzan and M.~Rajaee,
	arXiv:2007.14421 [hep-ph].
	
	
	\bibitem{Inelastic}
	K.~Harigaya, Y.~Nakai and M.~Suzuki,
	arXiv:2006.11938 [hep-ph];
	%
	H.~M.~Lee,
	arXiv:2006.13183 [hep-ph];
	%
	M.\ Baryakhtar, A.\ Berlin, H.\ Liu and N.\ Weiner,
	arXiv:2006.13918 [hep-ph];
	%
	J.~Bramante and N.~Song,
	arXiv:2006.14089 [hep-ph];
	%
	I.~M.~Bloch, A.~Caputo, R.~Essig, D.~Redigolo, M.~Sholapurkar, and T.~Volansky,
	arXiv:2006.14521 [hep-ph];
	%
	H.\,An and D.\,Yang, arXiv:2006. 15672 [hep-ph];
	%
	S.\,Baek, J.\,Kim, and P.\,Ko,
	arXiv:2006. 16876 [hep-ph];
	%
	A.~Aboubrahim, M.~Klasen and P.~Nath,
	arXiv:2011.08053 [hep-ph];
	%
	D.~Borah, S.~Mahapatra and N.~Sahu,
	[arXiv:2009.06294 [hep-ph]].
	
	
	\bibitem{He:2020wjs}
	H.~J.~He, Y.~C.~Wang, and J.~Zheng,
	JCAP {2101} (2021) 042
	[arXiv:2007.04963].
		

\bibitem{Gu:2018kmv}
E.g., P.~H.~Gu and H.~J.~He,
Phys. Rev. D\,99 (2019) 015025, no.1 
[arXiv:1808.09377];
%
N.~Okada and O.~Seto,
Phys. Rev. D\,101 (2020) 023522, no.2
[arXiv:1908.09277].

\bibitem{Gherghetta:2019coi}
T.\ Gherghetta, J.\ Kersten, K.\ Olive, and M.\ Pospelov,
Phys.\ Rev.\ D\,{100} (2019) 095001, no.9
[arXiv:1909.00696 [hep-ph]].

\bibitem{Aprile:2018dbl}
E.~Aprile \textit{et al.} [XENON Collaboration],
Phys.\ Rev.\ Lett.\\  {121} (2018) 111302 
[arXiv:1805.12562\,[astro-ph.CO]].


\bibitem{DarkS-2015}
P.\ Agnes {\it et al.} [DarkSide Collaboration],
Phys.\ Rev.\ D\,93 (2016) 081101, no.8 [arXiv:1510.00702 [astro-ph.CO]].

\bibitem{Migdal}
	A.\ B.\ Migdal, J.\ Phys.\,(USSR) 4 (1941) 449;
	M.\ Ibe, W.\ Nakano, Y.\ Shoji, and K.\ Suzuki,
	JHEP {03} (2018) 194
	[arXiv:1707.07258 [hep-ph]];
	M.\ J.\ Dolan, F.\ Kahlhoefer, C.\ McCabe,
	Phys.\ Rev.\ Lett.\ {121} (2018) 101801, 
    [arXiv:\ 1711.09906 [hep-ph]].


	\bibitem{Aprile:2019jmx}
	E.~Aprile \textit{et al.} [XENON Collaboration],
	Phys.\ Rev.\ Lett.\ {123} (2019) 241803, no.24
	[arXiv:1907.12771 [hep-ex]].


	\bibitem{Bell:2021zkr}
	N.~F.~Bell, J.~B.~Dent, B.~Dutta, S.~Ghosh, J.~Kumar, and J.~L.~Newstead,
	Phys.\ Rev.\ D {104} (2021) 7, no.7 
	[arXiv: 2103.05890 [hep-ph]].


    \bibitem{DarkS}
	P.~Agnes \textit{et al.} [DarkSide Collaboration],
	Phys.\ Rev.\ Lett.\ 121 (2018) 081307, no.8
	[arXiv:1802.06994 [astro-ph.\ HE]].
	

	\bibitem{CDEX-1B}
	Z.~Z.~Liu \textit{et al.} [CDEX Collaboration],
	Phys.\ Rev.\ Lett.\ 123 (2019) 161301, no.16
	[arXiv:1905.00354 [hep-ex]].


	\bibitem{Abdelhameed:2019hmk}
	A.~H.~Abdelhameed \textit{et al.} [CRESST Collaboration],
	Phys.\ Rev.\ D {100} (2019) 102002, no.10
	[arXiv:1904.00498 [astro-ph.CO]].


	\bibitem{Belanger:2008sj}
	G.\ Belanger, F.\ Boudjema, A.\ Pukhov, and A.\ Semenov,
	Comput.\ Phys.\ Commun.\ {180} (2009) 747
	[arXiv: 0803.2360 [hep-ph]].
	
	\bibitem{Yu:2011by}
	Z.\ H.\ Yu, J.\ M.\ Zheng, X.\ J.\ Bi, Z.\ Li, D.\ X.\ Yao, and H.\ H.\ Zhang,
	Nucl.\ Phys.\ B {860} (2012) 115
	[arXiv:1112.6052 [hep-ph]].
	
	
	\bibitem{Jackson:2013pjq}
	C.\ B.\ Jackson, G.\ Servant, G.\ Shaughnessy, T.~Tait, and M.\ Taoso,
	JCAP {07} (2013) 021 [arXiv:1302.1802].
	
	\bibitem{xray_DDM}
	R.~Essig, E.~Kuflik, S.~D.~McDermott, T.~Volansky, 
    and K.~M.~Zurek,
	JHEP \textbf{11}, 193 (2013)
	[arXiv:1309.4091 [hep-ph]];
	O.~Ruchayskiy, A.~Boyarsky, D.~Iakubovskyi, E.~Bulbul, D.~Eckert, J.~Franse, D.~Malyshev, M.~Markevitch and A.~Neronov,
	Mon.\ Not.\ Roy.\ Astron.\ Soc.\ 
    {460}, no.2, 1390-1398 (2016)
	[arXiv:1512.07217 [astro-ph.HE]].

	
	\bibitem{CMB_DDM}
	T.~R.~Slatyer and C.~L.~Wu,
	Phys.\ Rev.\ D {95} (2017) 023010, no.2 
	[arXiv:1610.06933 [astro-ph.CO]];
	L.~Zhang, X.~Chen, M.~Kamionkowski, Z.\,G.~Si and Z.~Zheng,
	Phys. Rev. D\,{76} (2007) 061301 
	[arXiv:0704.2444 [astro-ph]].
	
	\bibitem{Belanger:2020gnr}
	G.\,Belanger, A.\,Mjallal and A.\,Pukhov,
	arXiv:2003.08621 [hep-ph].
	
	\bibitem{PDG}
	M.~Tanabashi \textit{et al.} [Particle Data Group],
	Phys.\ Rev.\ D\,98 (2018) 030001.
	
	\bibitem{Leveille:1977rc}
	J.~P.~Leveille,
	Nucl.\ Phys.\ B {137} (1978) 63.
	
	
	\bibitem{Fox}
	P.~J.~Fox, R.~Harnik, J.~Kopp and Y.~Tsai,
	Phys.\ Rev.\ D {84} (2011) 014028
	[arXiv:1103.0240 [hep-ph]].
	
	
	\bibitem{Essig:2013vha}
	R.\ Essig, J.\ Mardon, M.\ Papucci, T.\ Volansky, and Y.\ M.\ Zhong,
	JHEP {11} (2013) 167
	[arXiv:1309.5084 [hep-ph]].
	
	
	\bibitem{Darme:2020ral}
	L.\ Darmé, S.\,A.\,R.\ Ellis, and T.\ You,
	JHEP {07} (2020) 053
	[arXiv:2001.01490 [hep-ph]].
	
	
	\bibitem{LEP_l_scatt}
	S.~Schael {\it et al.} [ALEPH, DELPHI, L3, OPAL and LEP Electroweak Collaborations],
	Phys.\ Rept.\  {532} (2013) 119
	[arXiv:1302.3415 [hep-ex]].
	
	
	\bibitem{DELPHI}
	J.~Abdallah {\it et al.} [DELPHI Collaboration],
	Eur.\ Phys.\ J.\ C {38} (2005) 395
	[hep-ex/0406019];
	J.~Abdallah {\it et al.} [DELPHI Collaboration],
	Eur.\ Phys.\ J.\ C {60} (2009) 17
	[arXiv:0901.4486 [hep-ex]].
	
	
	\bibitem{Aubert:2008as}
	B.~Aubert \textit{et al.} [BaBar Collaboration],
    arXiv:0808.0017 [hep-ex],
    presented at ICHEP-2008, Philadelphia, USA.
	
	
	\bibitem{Abe:2010gxa}
	T.~Abe \textit{et al.} [Belle-II Collaboration],
	Belle II Technical Design Report,
	arXiv:1011.0352 [physics.ins-det].
	
	
	\bibitem{LHC}
	M.~Aaboud {\it et al.} [ATLAS Collaboration],
	JHEP {1806} (2018) 166
	[arXiv:1802.03388 [hep-ex]];
	A.~M.~Sirunyan {\it et al.} [CMS Collaboration],
	Phys.\ Lett.\ B {792} (2019) 345
	[arXiv:1808.03684 [hep-ex]].
	
	
	\bibitem{ATLAS:2020wzf}
	[ATLAS Collaboration], ATLAS-CONF-2020-048.
	
	
	\bibitem{CMS}
	A.~M.~Sirunyan {\it et al.}
	[CMS Collaboration],
	Phys.\ Lett.\ B {792} (2019) 345
	[arXiv:1808.03684 [hep-ex]].
	
	
	\bibitem{Buchmueller:2014yoa}
	O.\ Buchmueller, M.\ J.\ Dolan, S.\ A.\ Malik, and C.\ McCabe,
	JHEP {01} (2015) 037
	[arXiv:1407.8257 [hep-ph]].
	
	
	\bibitem{PandaX4T}
	H.~Zhang {\it et al.} [PandaX Collaboration],
	Science China (Phys.\ Mech.\ Astron.) 62 (2019) 31011, no.3
	[arXiv: 1806.02229 [physics.ins-det]].
	
	
	\bibitem{LZ}
	D.\ Akerib {\it et al.} [LUX-ZEPLIN (LZ) Collaboration],
	Nucl.\ Instru.\ \& Meth.\ A\,953 (2020) 163047
	[arXiv: 1910.09124 [physics.ins-det]].
	
	
	\bibitem{XENONnT}
	E.\ Aprile {\it et al.} [XENON Collaboration],
	JCAP 2011 (2020) 031 [arXiv:2007.08796 [physics.ins-det]];
	%
	JCAP 1604 (2016) 027 [arXiv:1512.07501 [physics.ins-det]].
	

	\bibitem{2HDM}
	For a review of the 2HDM,
	G.~C.~Branco, P.~M.~Ferreira, L.~Lavoura, M.~N.~Rebelo, M.~Sher, and J.~P.~Silva,
	Phys.\ Rept.\ {516} (2012) 1 [arXiv:1106.0034 [hep-ph]]; and references therein.
	
	
	\bibitem{Cheng:1987rs}
	T.~P.~Cheng and M.~Sher,
	Phys.\ Rev.\ D {35} (1987) 3484
	
	
	\bibitem{Alpigiani:2017lpj}
	C.\ Alpigiani, A.\ Bevan, M.\ Bona, M.\ Ciuchini, D.\ Derkach,
	E.\ Franco, V.\ Lubicz, G.\ Martinelli, F.\ Parodi, and M.~Pierini \textit{et al.,}
	arXiv:1710.09644 [hep-ph].
	
	
	\bibitem{Aad:2019ojw}
	G.~Aad \textit{et al.} [ATLAS Collaboration],
	Phys.\ Lett.\ B {801} (2020) 135148
	[arXiv:1909.10235 [hep-ex]].
	
\end{thebibliography}
\end{document}